\documentclass[lettersize,journal]{IEEEtran}
\pdfoutput=1
\IEEEoverridecommandlockouts
\usepackage{amsmath,amsfonts}
\usepackage{algorithmic}
\usepackage{algorithm}
\usepackage{svg} % 引入 SVG 宏包
\usepackage{enumitem}
\usepackage{array}
\usepackage[caption=false,font=normalsize,labelfont=sf,textfont=sf]{subfig}
\usepackage{textcomp}
\usepackage[colorlinks=true, linkcolor=blue, citecolor=blue, urlcolor=blue]{hyperref}
\usepackage{stfloats}
\usepackage{url}
\usepackage{verbatim}
\usepackage{graphicx}
\usepackage{cite}
\makeatletter

\newcommand{\Rmnum}[1]{\expandafter\@slowromancap\romannumeral #1@}
\makeatother
\begin{document}

\title{Beam Squint Assisted Joint Angle-Distance \\Localization for Near-Field Communications}

\author{Aibiao Zhang, Weizheng Zhang, and Chiya Zhang,~\IEEEmembership{Member,~IEEE }
        % <-this % stops a space

\thanks{The authors are with the School of Information Science and Technology, Harbin Institute of Technology, Shenzhen 518055, China (e-mail: 24s152040@stu.hit.edu.cn).}}

\maketitle

\begin{abstract}
With the advent of extremely large-scale MIMO (XL-MIMO), mmWave/THz bands and ultra-wideband transmission, future 6G systems demand real-time positioning with centimeter or even millimeter level accuracy. This paper addresses the pronounced near-field beam squint problem caused by phase shifter based beamforming in wideband near-field scenarios and proposes a beam squint assisted joint angle–distance localization scheme. The key idea is to employ true-time-delay (TTD) units together with phase shifters (PS) to synthesize a controllable joint angle-distance (JAD) trajectory that establishes a unique mapping between subcarriers and spatial locations, enabling single scan acquisition of target angle and range. To implement this paradigm efficiently, we design a coarse to fine two stage estimator: a low complexity coarse stage based on subcarrier power peaks for user separation and candidate region selection, followed by a local high resolution refinement stage that applies spatial smoothing and near-field multiple signal classification (MUSIC) over multiple subcarriers and fuses the resulting spectra by geometric averaging to suppress spurious peaks. We theoretically prove the correctness and uniqueness of the MUSIC spatial spectrum peak under the proposed near-field steering model, and derive the Cramér–Rao lower bound (CRLB) for joint angle–distance estimation. Simulation results in single  and multi-user scenarios validate that the proposed method achieves very high accuracy and robustness, significantly outperforming conventional two-step approaches, and is promising for practical 6G sensing and localization deployments. 
\end{abstract}


\section{Introduction}
\IEEEPARstart{W}{ith} the commercialization of fifth-generation (5G) networks, global research has shifted toward sixth-generation (6G) systems \cite{9719949}. Compared with 5G, 6G targets more stringent requirements, including centimeter- or even millimeter-level positioning accuracy to enable seamless localization in both indoor and outdoor scenarios. Promising enablers include extremely large-scale multiple-input multiple-output (XL-MIMO) systems \cite{cui2022near1}, millimeter-wave/terahertz (mmWave/THz) communications \cite{rappaport2013millimeter}, and ultra-wideband transmission. Unlike conventional far-field massive MIMO (mMIMO) systems, where electromagnetic (EM) propagation is typically modeled by plane waves, XL-MIMO especially in high-frequency bands operates in the near-field region. This shift necessitates spherical wave channel modeling rather than traditional plane wave assumptions. The spherical wave characteristics allow beamforming to focus energy on specific spatial locations, offering strong potential for high precision user localization \cite{10750315}, \cite{10843386}, \cite{abu2021near}.

When users are located in the far-field region of the base station, early studies mainly focused on angle estimation, which can be formulated as a direction of arrival (DOA) estimation problem. In array signal processing, a number of classical DOA estimation techniques have been developed. For instance, the multiple signal classification (MUSIC) algorithm, which exploits eigen-decomposition of the covariance matrix, serves as a cornerstone in spatial spectrum estimation and direction finding \cite{schmidt1986multiple}, \cite{lin2006fsf}. In \cite{9068230}, an ESPRIT-based method with frequency scanning leaky wave antennas (LWA) was proposed to effectively estimate wideband DOA. For distributed source scenarios, a novel DOA estimation approach based on the minimum variance distortionless response (MVDR) criterion was introduced in \cite{10005294}, aiming to meet the requirements of real-time applications. More recently, Zhang et al. proposed a deep learning based DOA estimation scheme, where convolutional neural networks (CNNs) were employed to extract spatial features from received signals, thereby enabling high-accuracy DOA estimation \cite{10891008}.

While far-field studies have primarily focused on DOA estimation, near-field scenarios present fundamentally different challenges and opportunities. In the near-field region, the user’s distance information becomes equally critical. When users are located within this region, electromagnetic (EM) waves must be accurately modeled as spherical waves \cite{cui2022channel}, \cite{zhang2022beam}. The boundary between the far-field and near-field regions is determined by the Rayleigh distance, which is proportional to both the square of the array aperture and the carrier frequency. With the enlarged aperture and increased frequency in broadband XL-MIMO systems, the near-field region can extend to several tens or even hundreds of meters \cite{zhang2022beam}. Consequently, in XL-MIMO, beam focusing should be achieved by exploiting spherical wave characteristics to concentrate signals on specific spatial positions, rather than merely steering them toward angular directions as in conventional far-field beamforming \cite{liu2024near}, \cite{liu2025near}, \cite{11030222}, \cite{khamidullina2021conditional}, \cite{guerra2021near}. For example, Hua et al. investigated integrated sensing and communication (ISAC) systems in the near-field, where echo signals were simultaneously utilized for both communication and localization \cite{11030222}. In \cite{khamidullina2021conditional}, the performance of several tensor decomposition based three dimensional (3D) localization estimators was compared. Furthermore, a Bayesian tracking algorithm was developed in \cite{guerra2021near} to track a single near-field target in 3D space.

\IEEEpubidadjcol

\subsection{Related Works}
However, extremely large bandwidths introduce the beam squint effect, where beams at different frequencies focus on different spatial locations. As a result, many frequency-dependent beams may not align with the true user position, leading to significant array gain loss. Since the number of antennas in current 5G massive MIMO systems remains relatively limited \cite{10737444}, existing studies have mainly focused on far-field beam squint issues. A large body of work has demonstrated that the beam squint effect can be exploited by carefully designing true-time-delay (TTD) and phase shifter (PS) parameters, such that frequency-dependent beams scan multiple directions simultaneously. This property enables efficient beam training and tracking \cite{luo2024yolo}, \cite{10058989}, \cite{10886954}, \cite{zhou2025radar}. For instance, Luo et al. \cite{luo2024yolo} adjusted PS and TTD values to control the squinting range, establishing a one-to-one mapping between subcarrier frequencies and angles, so that user directions can be estimated from the peak frequency of the received echoes. In \cite{10058989}, beam squint was leveraged to rapidly acquire user angles in an integrated sensing and communication (ISAC) system. In \cite{10886954}, the squint effect was exploited for joint communication and localization in hybrid-field scenarios, where a novel hybrid beamformer was designed by combining beam broadening and focusing. Moreover, Zhou et al. \cite{zhou2025radar} proposed a frequency-dependent “rainbow beam” design for ISAC, where a single OFDM symbol suffices to accomplish fast beam training by exploiting beam squint.

For XL-MIMO systems, the extremely large array aperture, the massive number of antennas, and the wide bandwidth at high frequencies give rise to a pronounced near-field beam squint effect. Hence, more practical studies should account for near-field squint rather than relying solely on far-field assumptions. Several recent works have explicitly investigated this phenomenon \cite{cui2022near2}, \cite{zheng2024near}, \cite{luo2023beam}, \cite{lei2024deep}. In \cite{cui2022near2} and \cite{zheng2024near}, it was shown that TTD can be employed to control the degree of near-field squint. By leveraging TTD-based beamforming, multiple beams focusing at different positions across frequencies can be generated, thereby enabling efficient near-field beam training with reduced training overhead. In \cite{luo2023beam}, the authors jointly optimized PS and TTD parameters to realize controllable squint beams and derived the trajectory equation of near-field squint points. This mapping ensures that users located at different positions receive peak power on different subcarriers, which can be directly exploited for angle–distance localization. Building upon \cite{luo2023beam}, Lei et al. \cite{lei2024deep} further considered the spatial non-stationarity inherent in near-field scenarios and proposed a CNN based localization framework that integrates controllable squint beams with beam training.

\subsection{Motivations and Contributions}
However, most existing near-field localization methods adopt a two stage strategy, where the user angle is first estimated via horizontal beam scanning, followed by distance estimation through radial beam scanning. Such a sequential procedure is prone to error propagation: estimation errors in the angular domain are further amplified during subsequent distance estimation, thereby degrading overall localization accuracy, sensing reliability, and latency performance. This limitation makes it difficult to satisfy the stringent requirements of ultra reliable low latency communications (URLLC). To the best of the authors’ knowledge, no prior work has investigated the joint estimation of angle and distance for user localization in the near-field.

To fill this gap, we propose a beam squint assisted MUSIC scheme for joint angle–distance localization in the near-field, which enables the base station to acquire user positions through a single scanning process. Our main contributions are summarized as follows:
  \begin{itemize}
    \item We propose an innovative “single-scan” near-field localization paradigm enabled by the joint use of TTD units and PS. By designing controllable two-dimensional beam trajectories under wideband conditions, a one-to-one mapping between subcarriers and the angle–distance parameters of the target is established, allowing the base station to obtain two dimensional user positions within a single frequency sweep. This approach effectively eliminates the error propagation and latency accumulation inherent in conventional two-step localization schemes. Furthermore, we theoretically prove the uniqueness of the near-field MUSIC spectrum under the proposed trajectory design, ensuring the stability and feasibility of the algorithm. This new paradigm offers a promising solution to meet the stringent requirements of URLLC in future 6G networks.
    \item We develop a two stage coarse-to-fine estimation algorithm to efficiently implement the proposed paradigm. In the first stage, a coarse subcarrier index is identified from the received power spectrum peak and combined with the trajectory relation to yield an initial angle–distance estimate. In the second stage, a near-field steering vector is constructed in the neighborhood of the coarse estimate, where spatial smoothing and eigenvalue decomposition are applied to extract the noise subspace. A local MUSIC spectrum search then provides refined estimates. Finally, multi subcarrier MUSIC spectra are fused via geometric averaging to suppress spurious peaks and enhance robustness. This algorithm significantly reduces computational complexity while maintaining stable performance across different SNRs and array configurations. In addition, we derive the Cramér–Rao lower bound (CRLB) as a theoretical benchmark.
    \item We conduct extensive simulations and comparative experiments to validate the proposed method. Results show that, compared with conventional two step schemes and power peak baselines, our approach achieves superior localization accuracy within a single scan, with angle root mean square error (RMSE) below 0.001° and distance RMSE below 0.001m. The method also demonstrates strong two dimensional resolution in both single user and multi user scenarios, and remains robust even under severe near-field beam-squint effects in broadband systems. The proposed scheme simultaneously achieves low latency, high accuracy, and high reliability, highlighting its strong potential for deployment in future 6G ultra-dense networks and intelligent sensing applications.
  \end{itemize}

\subsection{Organization and Notation}
The remainder of this paper is organized as follows. Section \ref{sec:system_model} introduces the system model and near-field channel modeling. Section \ref{sec:beam_squint} analyzes the near-field beam squint phenomenon and reviews the controllable beam squint strategy enabled by TTDs. In Section \ref{sec:localization}, we propose a joint angle distance localization scheme based on the MUSIC algorithm with beam squint assistance. Section \ref{sec:simulation} and Section \ref{sec:conclusion} present the simulation results and discussions, respectively.
  
The notations used throughout this paper are summarized as follows. Scalars are denoted by italic letters, while vectors and matrices are represented by bold lowercase and bold uppercase letters, respectively. For a matrix $\mathbf{A}$, $\mathbf{A}^T$, $\mathbf{A}^H$, and $\mathbf{A}^{-1}$ denote its transpose, conjugate transpose, and inverse, respectively. The notation $|\cdot|$ represents the absolute value of a scalar or the determinant of a matrix, while $|\cdot|$ denotes the Euclidean norm of a vector. The operator $\mathbb{E}{\cdot}$ stands for statistical expectation, and $\Re{\cdot}$ denotes the real part. The symbols $ \otimes $, $ \odot $, and $ \circ $ denote the Kronecker product, Hadamard product, and outer product of vectors, respectively. 

\section{System Model}
\label{sec:system_model}
We consider a broadband large-scale multiple input and multiple output (MIMO) system operating in the millimeter-wave (mmWave) band, where orthogonal frequency-division multiplexing (OFDM) modulation is adopted. The base station (BS) is equipped with a uniform linear array (ULA) consisting of $N$ antennas with inter element spacing $d$, and employs a single radio-frequency (RF) chain architecture. The $n$-th antenna element is located at coordinates $(0, nd)$, where $n=-(N-1)/2, \ldots, (N-1)/2$. The array aperture is $D=(N-1)d \approx Nd$. The carrier frequency and transmission bandwidth are denoted by $f_c$ and $W$, respectively, with subcarrier frequencies distributed over $[f_c-W/2, f_c+W/2]$. A total of $M+1$ subcarriers are assumed, where the 0-th subcarrier is $f_0=f_c-W/2$, and the frequency of the $m$-th subcarrier is given by$f_m=f_0+\widetilde{f}_m$.Let $\widetilde{f}_m = m\cdot (W/M)$ denote the baseband frequency of the $m$-th subcarrier, so that $f_m=f_0+\widetilde{f}_m$, where $m=0,1,2,\ldots,M$. The BS serves $K$ single-antenna users. The $k$-th user is located at Cartesian coordinates $(x_k,y_k)$, which correspond to polar coordinates $(r_k,\theta_k)$.

\begin{figure}[!htbp]
\centering
\hspace{-1cm} % 向左移动
\includegraphics[width=2.5in]{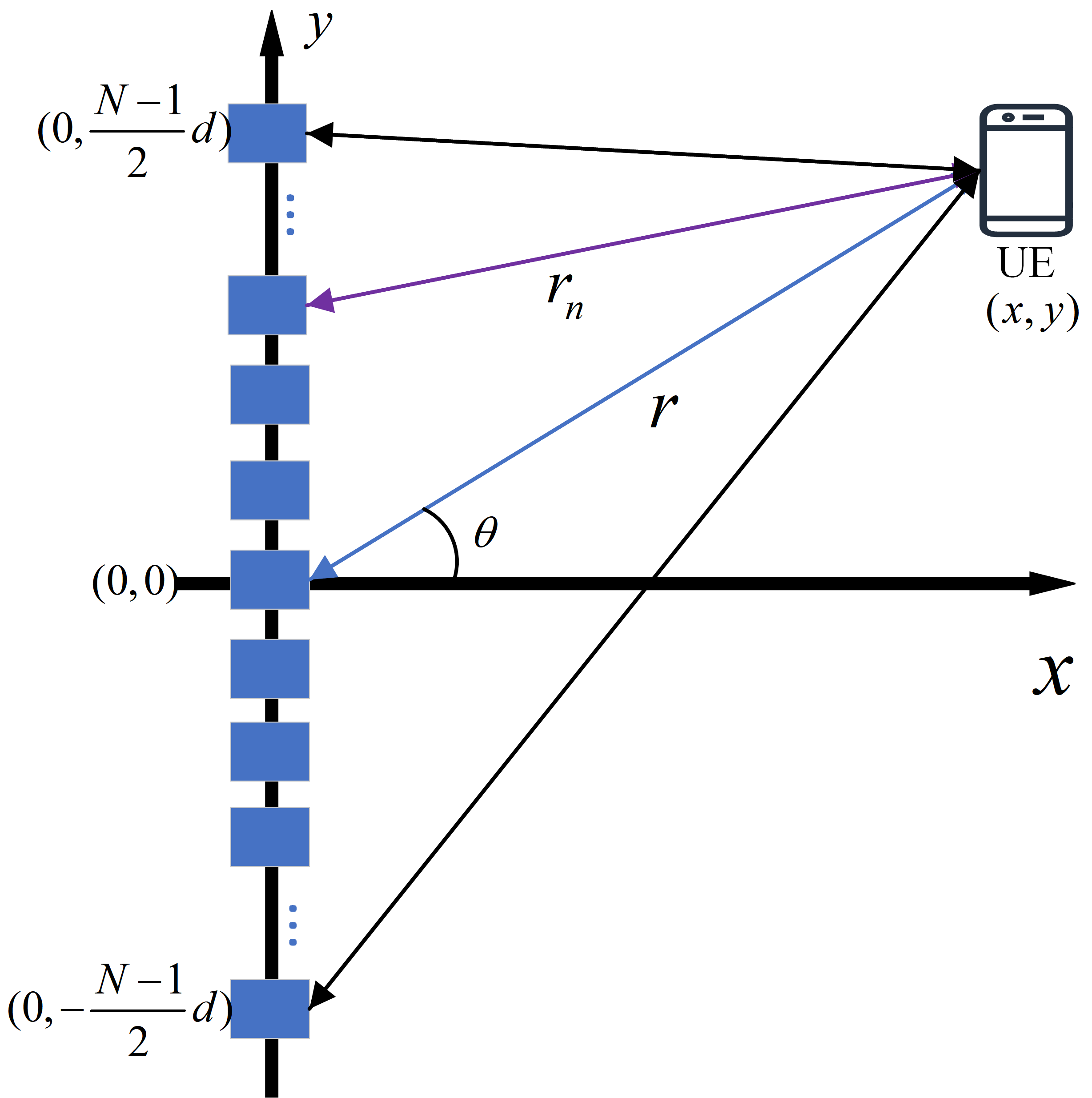}
\caption{Near-field system model.}
\label{model}
\end{figure}

Due to severe path loss caused by scattering, millimeter-wave and terahertz communications rely heavily on line-of-sight (LoS) links. Therefore, we primarily focus on near-field LoS channels,although the discussion in this paper can be directly extended to non-line-of-sight (NLoS) scenarios \cite{elayan2019terahertz}. According to classical antenna theory \cite{dandekar2007reviews}, the boundary between the near-field and far-field regions is determined by the Rayleigh distance $Z=2D^2/\lambda$, where $\lambda=c/f_c$ denotes the wavelength at the carrier frequency. Let $r_{k,n}$ represent the distance between the $n$-th BS antenna and the $k$-th user antenna. The corresponding propagation delay $\tau_{k,n}$ is given by $\tau_{k,n}=\frac{r_{k,n}}{c}$,where $c$ is the speed of light. In the near-field region, $r_{k,n}$ is commonly expanded using the Fresnel approximation as follows:
		\begin{equation}
		\begin{aligned}
			r_{k,n} &= \sqrt{x_k^2 + (y_k - n d)^2} 
			= \sqrt{r_k^2 + n^2 d^2 - 2 r_k n d \sin\theta_k} \\
			&\approx r_k - n d \sin\theta_k + \frac{n^2 d^2 \cos^2\theta_k}{2 r_k}
		\end{aligned}
	\end{equation}

Under the spherical wavefront assumption, the frequency domain channel between the $n$-th BS antenna and the $k$-th UE at the $m$-th subcarrier can be expressed as $h_{k,n,m}=\beta_{k,n,m}e^{-j2\pi f_m\tau_{k,n}}$.According to standard channel modeling, the channel fading coefficient $\beta(k,n,m)$ is given by $\beta(k,n,m)=\frac{c}{4\pi f_m r_{k,n}}$.Since the user to BS center distance $r_k$ is much larger than the array aperture $D$ ($r_k \gg D$), it is commonly assumed that the path loss from all antennas to the user is approximately equal, i.e.,$\alpha_{k,-(N-1)/2,m}\approx\cdots\approx\alpha_{k,(N-1)/2,m}\approx\alpha_{k,m}=\frac{c}{4\pi f_m r_k}$.Therefore, the near-field LoS channel vector between the BS and the $k$-th user at the $m$-th subcarrier, denoted as $\mathbf{h}_{k,m} \in \mathbb{C}^{N \times 1}$, can be expressed as
\begin{equation}
	\mathbf{h}_{k,m} = \sqrt{N} \beta_{k,m} \mathbf{a}(r_k, \theta_k, f_m),
\end{equation}
where
\begin{equation}
\begin{array}{l}
  \mathbf{a}(r_k, \theta_k, f_m)\\
  =\frac{1}{\sqrt{N}} \left[ e^{-j\frac{2\pi f_m}{c} r_{k,\frac{-(N-1)}{2}}},  \ldots, e^{-j\frac{2\pi f_m}{c} r_{k,\frac{(N-1)}{2}}} \right]
\end{array}
\end{equation}
is the near-field array response vector.

To provide users with improved communication links, the BS employs single RF chain beamforming based on phase shifters (PS). Assume that the optimal near-field beamforming vector $\mathbf{w}(r_{0},\theta_{0}) \in \mathbb{C}^{N\times 1}$ is determined by the target position $(r_{0},\theta_{0})$ and the lowest subcarrier frequency $f_0$, i.e., $\mathbf{w}(r_{0},\theta_{0})$ should be conjugate-matched to the channel at $f_0$ corresponding to the reference position $(r_{0},\theta_{0})$ can be represented as
	\begin{equation}
		{{[w({{r}_{0}},{{\theta }_{0}})]}_{n}}=\frac{1}{\sqrt{N}}{{e}^{j\arg \left( {{[h({{r}_{0}},{{\theta }_{0}},{{f}_{0}})]}_{n}} \right)}}=\frac{1}{\sqrt{N}}{{e}^{-j2\pi {{f}_{0}}\frac{{{r}_{0,n}}}{c}}}
	\end{equation}

The downlink transmission process for user $k$ at subcarrier $m$ can be expressed as:
\begin{equation}
		y_{k,m}=\mathbf{h}_{k,m}^H\cdot\mathbf{w}\cdot s_{k,m}+n_{k,m},
	\end{equation}
where $y_{k,m} \in \mathbb{C}^{1 \times 1}$ denotes the complex signal received by the $k$-th user at the $m$-th subcarrier.And $\mathbf{h}_{k,m} \in \mathbb{C}^{N \times 1}$ denotes the near-field channel vector, $\mathbf{w} \in \mathbb{C}^{N \times 1}$ is the beamforming vector, $s_{k,m}$ is the transmitted signal at subcarrier $m$, and $n_{k,m}$ represents additive white Gaussian noise (AWGN).

\section{Near Field Controllable Beam Squint}
\label{sec:beam_squint}

In this section, we first analyze the beam squint effect in near-field communications. Then, we will briefly review the controllable beam squint control method proposed in \cite{luo2023beam}.

\subsection{Near-Field Beam Squint Effect}
To focus the transmitted beam of the BS on a desired near-field position, conventional beamforming is typically implemented using phase shifters. In narrowband systems, since the frequency differences among subcarriers are small, all subcarriers approximately focus on the desired location. However, when the bandwidth becomes large, the phase shifts introduced by phase shifters are independent of frequency, resulting in significant phase response differences across subcarriers. This frequency space coupling effect causes the main lobe of the same beam to deviate in different directions across subcarriers, thereby generating multiple beams with distinct shapes and locations in space. This phenomenon is referred to as near-field beam squint, which can lead to severe performance degradation in wideband XL-MIMO systems.

In wideband systems, beamforming implemented by phase shifters focuses on $M$ different positions that are frequency dependent, where $M$ denotes the number of subcarriers. When the beamforming vector $\mathbf{w}(r_0, \theta_0)$ is designed based on the desired location $(r_0, \theta_0)$ and the frequency $f_0$, the array gain achieved by the phase-shifter-based beamforming vector $\mathbf{w}(r_0, \theta_0)$ at an arbitrary user position $(r, \theta)$ can be expressed as

\begin{align}
    &g(r,\theta, f_{m},\mathbf{w}) = \left| \mathbf{w}^{H} \cdot \mathbf{a}(r,\theta, f_{m}) \right|  \notag\\
    &= \frac{1}{N} \left| \sum_{n=-\frac{N-1}{2}}^{\frac{N-1}{2}} e^{j2\pi f_{0}\frac{r_{0,n}}{c}} e^{-j2\pi f_{m}\frac{r_{n}}{c}} \right| \notag \\
    &= \frac{1}{N} \left| \sum_{n=-\frac{N-1}{2}}^{\frac{N-1}{2}} 
    \begin{aligned}
        &e^{j\frac{2\pi n^2 d^2}{c} \left(f_m \frac{\cos^2 \theta}{2r} - f_0 \frac{\cos^2 \theta_0}{2r_0}\right)} \\ % 在此处换行
        &\quad \times e^{-j\frac{2\pi nd}{c} \left(f_m \sin \theta - f_0 \sin \theta_0\right)}
    \end{aligned}
    \right|.
\end{align}

The above expression is relatively complex. To illustrate its structure more clearly, we define a function $G(x,y)$, which can be expressed as
\begin{equation}
\label{G_function}
	G(x, y) = \frac{1}{N} \left| \sum_{n=-N}^{N} \exp\left(j\frac{2\pi n^2 d^2}{c} x + j \frac{2\pi nd}{c} y\right) \right|
\end{equation}

Then, the array gain can be expressed as
\begin{equation}
\begin{aligned}
    &g(r,\theta, f_{m},\mathbf{w}) \\
    &= G\left(f_m \frac{\cos^2 \theta}{2r} - f_0 \frac{\cos^2 \theta_0}{2r_0}, f_0 \sin \theta_0 - f_m \sin \theta\right)
\end{aligned}
\end{equation}

To determine the beam-squint focusing position, it is necessary to compute the maximum of the array gain $g(r,\theta, f_{m},\mathbf{w})$. This problem can be equivalently transformed into maximizing the function $G(x,y)$. From the functional form of \eqref{G_function}, it follows that the maximum of $G(x,y)$ is achieved when $x=0$ and $y=0$, that is, when

\begin{align}
    f_m \frac{\cos^2 \theta}{2r} - f_0 \frac{\cos^2 \theta_0}{2r_0} &= 0,
\end{align}

\begin{align}
    f_0 \sin \theta_0 - f_m \sin \theta&= 0.
\end{align}

Then, a feasible solution for $(r,\theta)$ can be expressed as
\begin{equation}
\label{eq:solution}
    \sin \theta = \frac{f_0}{f_m} \sin \theta_0, 
\end{equation}
\begin{equation}
\label{eq:solution2}
	r = r_0 \cdot \frac{f_m}{f_0} \cdot \frac{\cos^2 \theta}{\cos^2 \theta_0}.
\end{equation}

\begin{figure}[!t]
\centering
\includegraphics[width=3in]{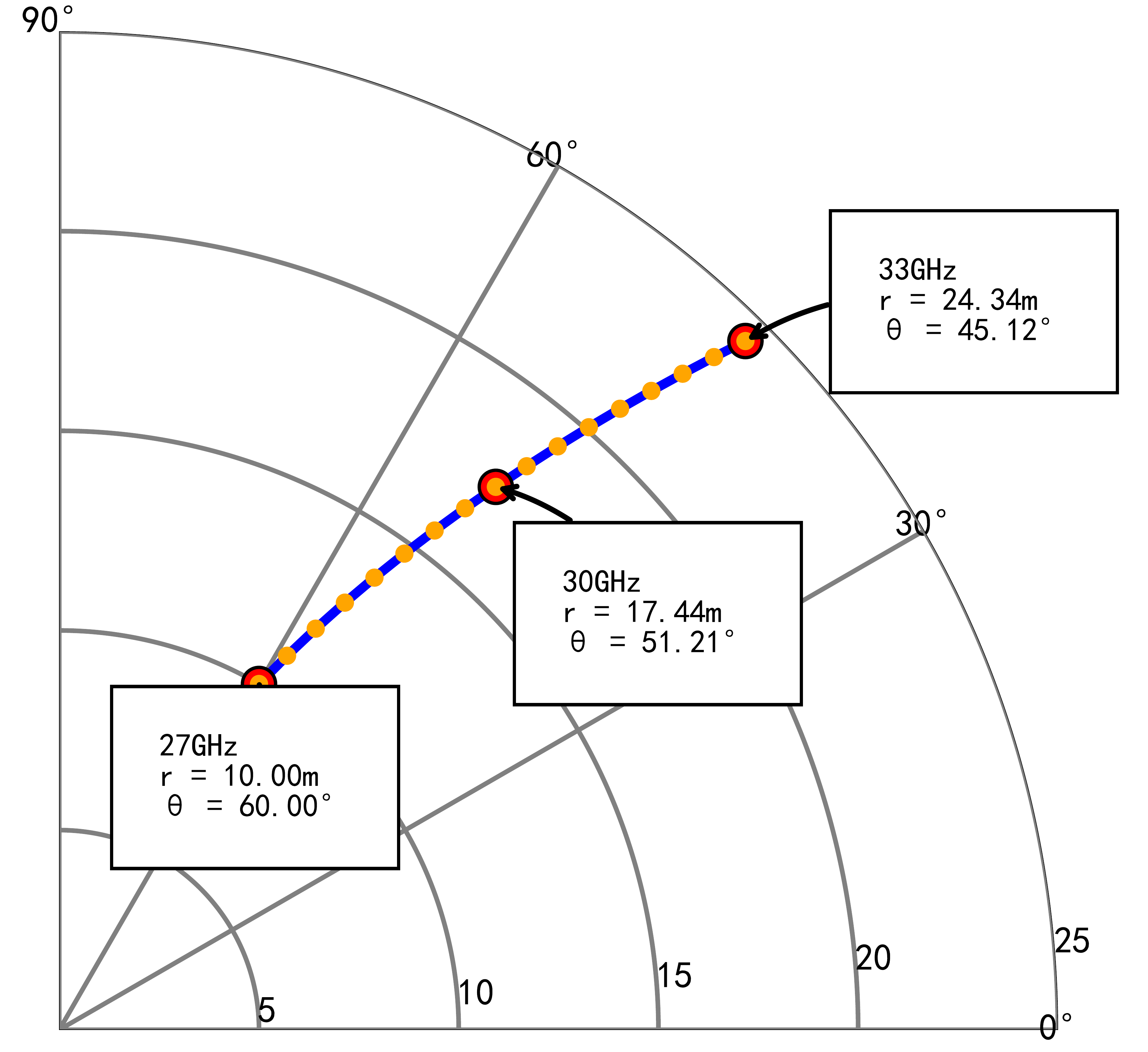}
\caption{Near-field beam squint}
\label{squint}
\end{figure}

Therefore, the near-field beam squint focusing point $(r_m,\theta_m)$ of the $m$-th subcarrier is determined by \eqref{eq:solution} and \eqref{eq:solution2}. Fig. \ref{squint} illustrates an example that clearly demonstrates the near-field beam-squint effect. As the subcarrier frequency $f_m$ increases, the beamforming focus shifts to different spatial locations across subcarriers, and these focusing points form a continuous trajectory. The trajectory starts from the focus of the 0-th subcarrier at frequency $f_0$ and ends at that of the $M$-th subcarrier at frequency $f_M$. In Fig. \ref{squint}, for instance, the beamforming focus at $f_0=27$ GHz is located at $(10\text{m}, 60^\circ)$, while the focus at $f_c=30$ GHz is $(17.44\text{m}, 51.22^\circ)$, and the highest frequency subcarrier at $f_M=33$ GHz is steered to $(24.34\text{m}, 45.12^\circ)$. These results clearly show that the near-field beam squint effect becomes significant in wideband systems and cannot be neglected.

\subsection{Controllable Beam Squint}

\begin{figure}[!t]
\centering
\hspace{-0.5cm} % 向左移动
\includegraphics[width=3in]{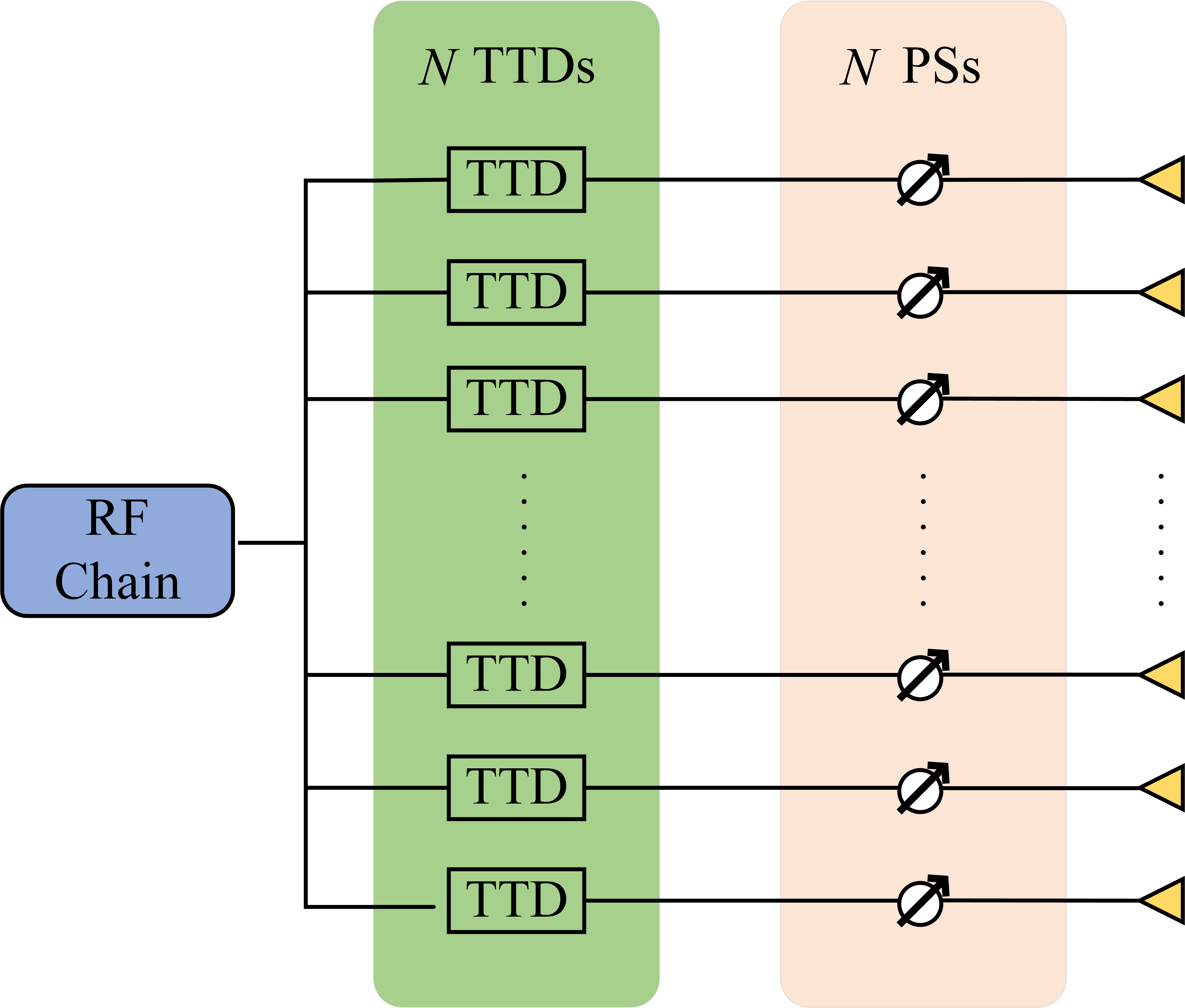}
\caption{TTD-PS beamforming architecture.}
\label{TTD}
\end{figure}

Recently, several studies have focused on the beam squint effect in near-field XL-MIMO systems. A feasible solution to this problem is to adopt beamforming based on TTD units rather than relying solely on PS. By leveraging TTDs, frequency dependent beams can be generated, thereby enabling controllable near-field beam squint. As illustrated in Fig. \ref{TTD}, assume that the BS is equipped with $N$ TTD lines, each cascaded with a phase shifter. Then, at the $m$-th subcarrier, the near-field beamforming vector $w(k,m) \in \mathbb{C}^{N\times 1}$ for the $k$-th user based on TTD can be rewritten as

\begin{figure*}[!t]
\centering
\subfloat[]{\includegraphics[width=2.5in]{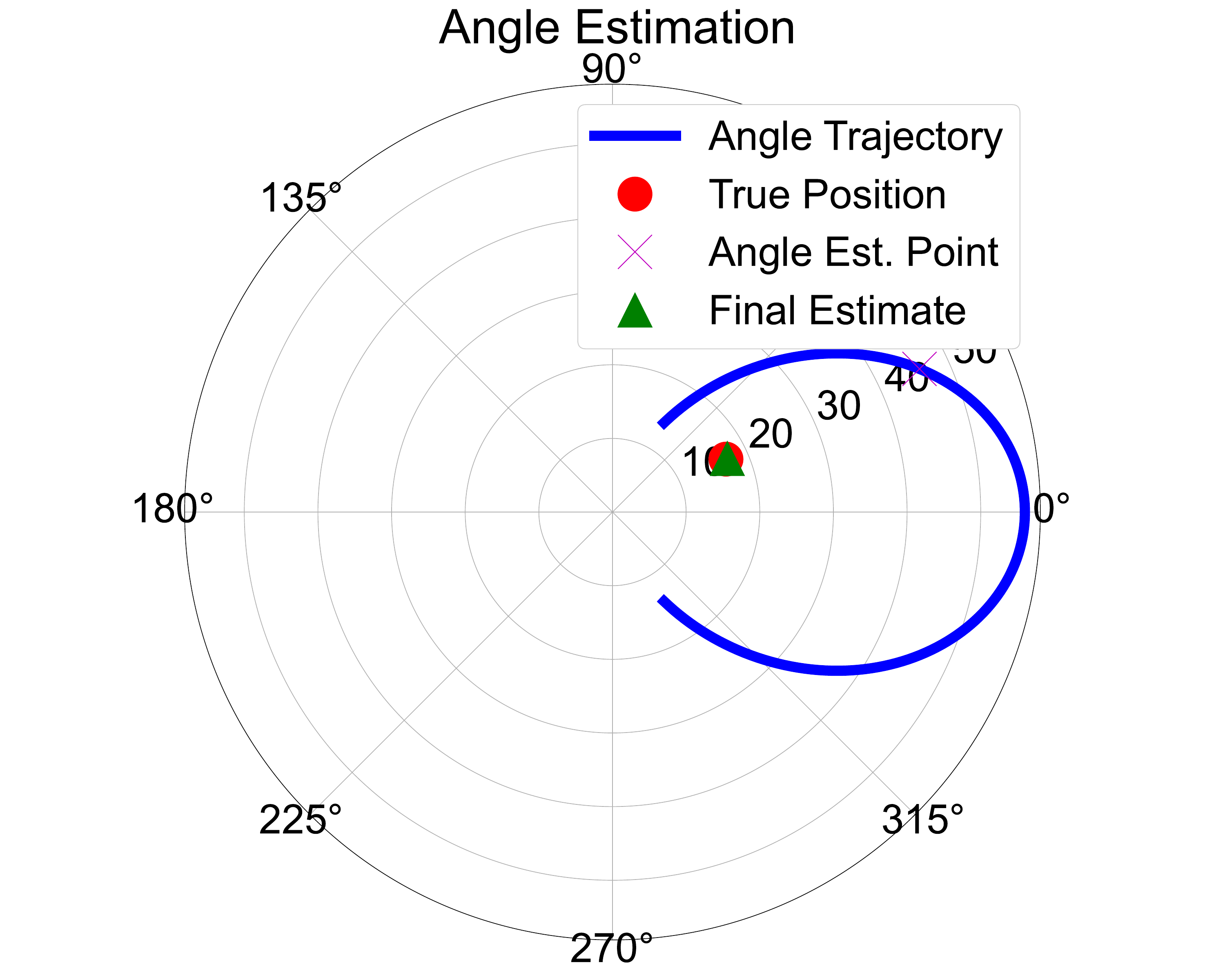}%
\label{T4}}
\hfil
\subfloat[]{\includegraphics[width=2.5in]{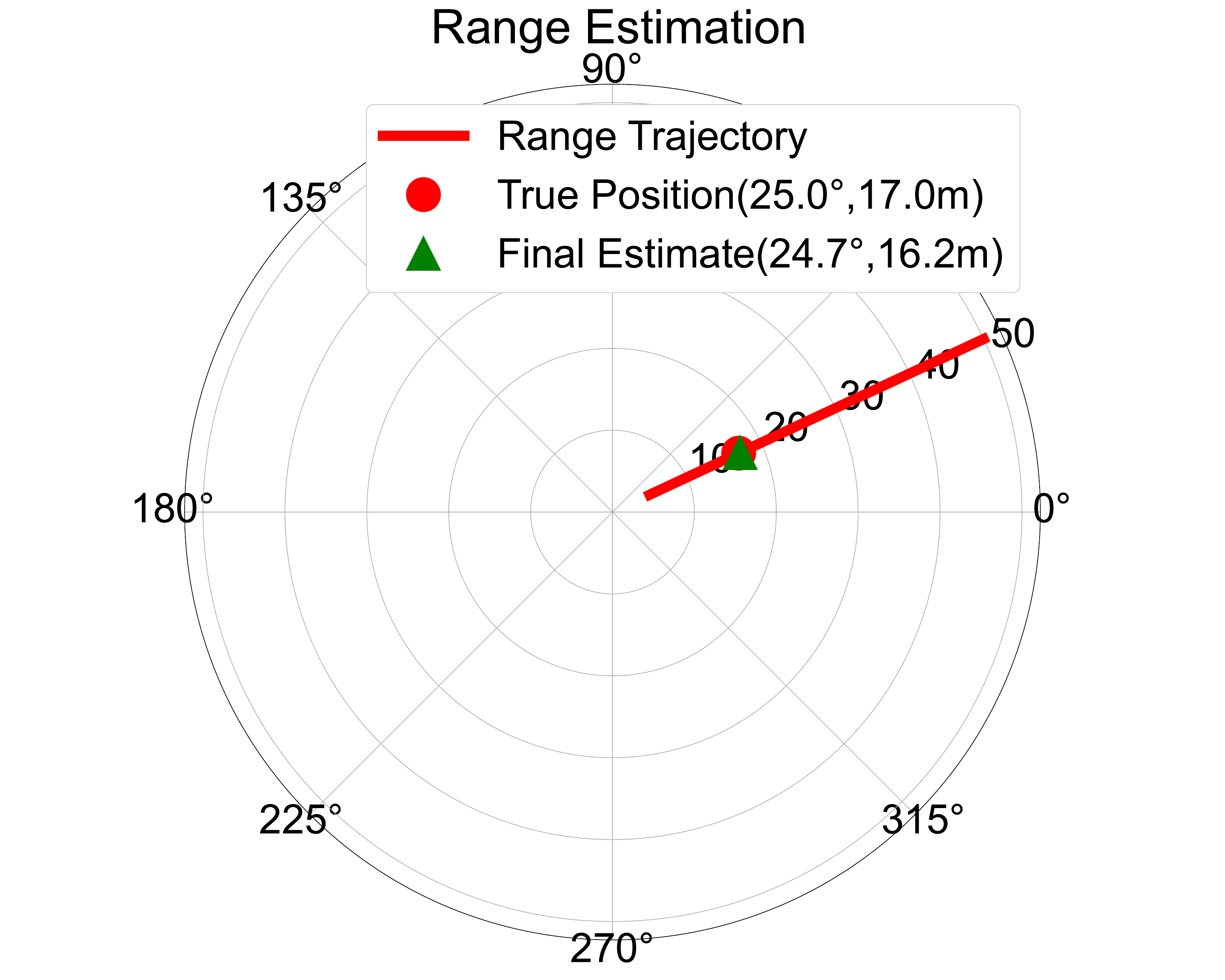}%
\label{T1}}
\caption{Illustration of the cascaded two step trajectory design reproduced from \cite{luo2023beam}. (a) Angular scanning trajectory for user angle estimation. (b) Radial scanning trajectory for range estimation with the angle fixed. This scheme requires two separate scans, and the angle estimation error propagates into the subsequent range estimation.}
\label{4}
\end{figure*}

\begin{equation}
	{{[\mathbf{w}]}_{n}}=\frac{1}{\sqrt{N}}{{e}^{-j2\pi {{\phi }_{n}}}}{{e}^{-j2\pi \tilde{f}_{m}{{t}_{n}}}}
\end{equation}
where $\phi_n$ and $t_n$ denote the phase shift and time delay of the $n$-th antenna, respectively. By jointly optimizing $\{\phi_n\}$ and $\{t_n\}$, the beamforming focus at different frequencies can be flexibly controlled.$\tilde{f}_m$ denotes the baseband frequency of the $m$-th subcarrier.

As shown in \cite{luo2023beam}, the near-field beam squint can be precisely controlled by adjusting the delays of the TTDs and the phase shifts of the PSs. By aligning the focus of the lowest subcarrier $f_0$ with the starting point $(\theta_s,r_s)$ and that of the highest subcarrier $f_M$ with the ending point $(\theta_e,r_e)$, the controllable focusing position corresponding to the maximum beam gain of the $m$-th subcarrier can be expressed as

\begin{equation}
		\sin \theta_m=\frac{\left(W-\tilde{f}_m\right) f_0}{W f_m} \sin \theta_s+\frac{\left(W+f_0\right) \tilde{f}_m}{W f_m} \sin \theta_e
		\label{theta}
\end{equation}
\begin{equation}
		\frac{1}{r_m}=\frac{1}{r_s} \frac{\left(W-\tilde{f}_m\right) f_0}{W f_m} \frac{\cos ^2 \theta_s}{\cos ^2 \theta_m}+\frac{1}{r_e} \frac{\left(W+f_0\right) \tilde{f}_m}{W f_m} \frac{\cos ^2 \theta_e}{\cos ^2 \theta_m} \
		\label{r}
\end{equation}

Assuming that the BS transmits an all-one pilot signal, the complex received signal of the $k$-th user at the $m$-th subcarrier under ideal noise-free conditions can be expressed as
	\begin{equation}
		\begin{split}
			y_{k,m} &= \mathbf{h}^H_{k,m} (r_k, \theta_k, f_m) \cdot \mathbf{w} \\
			&= \frac{\alpha_{k,m}}{\sqrt{N}} \sum_{n=-\frac{N-1}{2}}^{\frac{N-1}{2}} e^{j2\pi f_m \frac{r_{k,n}}{c} - j2\pi \phi_n - j2\pi \tilde{f}_m t_n}.
		\end{split}
	\end{equation}

In \cite{luo2023beam}, angle estimation is first performed by designing an angular scanning trajectory, and the user angle is obtained from the peak power, as shown in Fig. \ref{4} \subref{T4}. With the estimated angle fixed, a distance scanning trajectory is then constructed by tuning TTD and PS parameters, and the user distance is obtained from the corresponding peak power, as shown in Fig. \ref{4} \subref{T1}. This two-scan strategy suffers from error propagation, since inaccuracies in angle estimation are carried over to distance estimation, which degrades the overall localization performance.

\subsection{JAD Trajectory Design}

\begin{figure}[!t]
\centering
\includegraphics[width=2.1in]{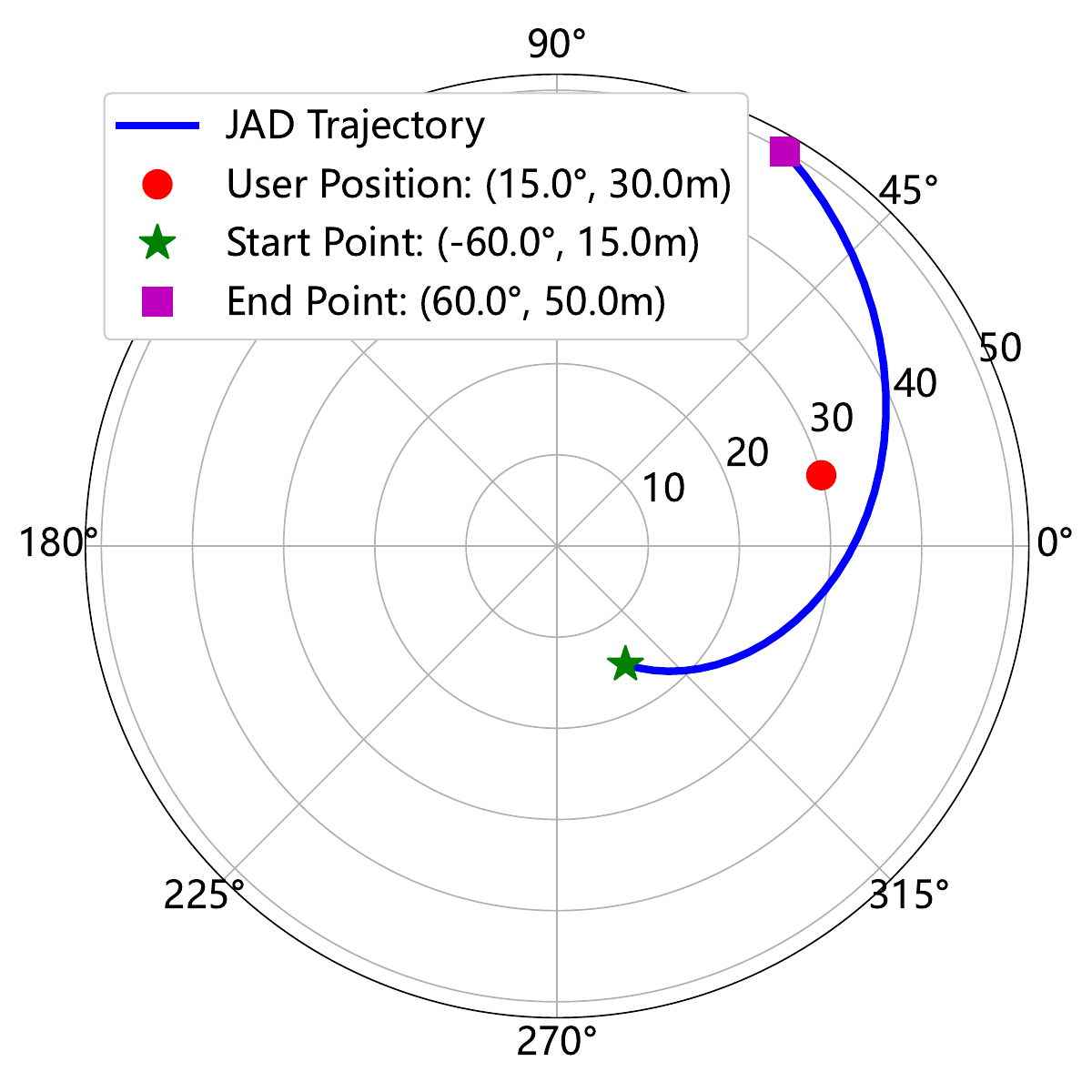}
\caption{Joint angle-distance trajectory. Where the trajectory squints from $(-60^\circ, 15\text{m})$ to $( 60^\circ, 50\text{m})$,the UE is located at $(15^\circ, 30\text{m})$.}
\label{JAD}
\end{figure}

 To overcome this limitation and achieve joint angle-distance (JAD) estimation, we propose a novel trajectory design that simultaneously optimizes the scanning paths for both angle and distance. First, we define the sensing region required by the BS system, where the user may exist within the minimum and maximum distances $r_{\text{min}}$ and $r_{\text{max}}$, and between the minimum and maximum angles $\theta_{\text{min}}$ and $\theta_{\text{max}}$, i.e.,$\{(\theta,r) |  \theta_{\text{min}} \le \theta \le \theta_{\text{max}},r_{\text{min}} \le r \le r_{\text{max}}\}$.A starting focusing point $(\theta_s,r_s)=(\theta_{\text{min}}, r_{\text{min}})$ and an ending focusing point $(\theta_e, r_e)=(\theta_{\text{max}}, r_{\text{max}})$ are then selected. These points should ensure that the beam-squint trajectory formed by all subcarriers can effectively cover the defined two-dimensional sensing region. For example, as illustrated by the JAD trajectory in Fig. \ref{JAD}, the trajectory squints from $(-60^\circ, 15\text{m})$ to $( 60^\circ, 50\text{m})$, thereby spanning a wide range of distances and angles. The next section we will presents a beam squint assisted MUSIC scheme that enables joint angle–distance estimation in a single scan, thereby eliminating error accumulation.

\section{User Localization}
\label{sec:localization}
In this section, we propose a “Coarse to Fine” two stage signal processing framework. The framework first performs a fast coarse localization by analyzing the power spectrum of the echo signal, achieving this with minimal computational cost. Then, using the coarse estimate as strong prior information, a fine grained two dimensional spectral peak search is conducted within a small local region. This is done using a near-field MUSIC subspace algorithm with super resolution capabilities, enabling joint estimation of both angle and distance in a single step.

\subsection{Coarse Localization via Power Spectrum}

By configuring the parameters of the TTDs and PSs according to the designed JAD trajectory, the received signal power of the $k$-th user at the $m$-th subcarrier $f_m$ can be expressed as
\begin{equation}
	P_{k,m} = |y_{k,m}|^2,
\end{equation}

If the user is located at $(r_k, \theta_k)$, its received power reaches the maximum at a specific subcarrier $f_{m^\star}$. We define the peak index as
\begin{equation}
	m^\star = \arg\max_{m} P_{k,m},
\end{equation}

Then, according to the trajectory design formulas \eqref{theta} and \eqref{r}, the coarse position estimate can be obtained as
\begin{equation}
	(\hat{r}_k^{(0)}, \hat{\theta}_k^{(0)}) = (\tilde{r}_{m^\star}, \tilde{\theta}_{m^\star}).
\end{equation}

This step does not require grid-based search, resulting in low computational complexity and making it suitable for fast coarse localization.

Although the coarse estimation algorithm is highly efficient, its performance is limited by two fundamental factors: (1) spatial quantization error, since the true target location usually lies between the discrete focusing points of the JAD trajectory; and (2) noise sensitivity, where low signal to noise ratio (SNR) conditions may cause the spectral peak to deviate. These limitations indicate that coarse estimation alone cannot meet the requirements of high-accuracy localization. Therefore, it is essential to introduce a super-resolution algorithm that can surpass the Rayleigh limit and fully exploit the phase information of the array.

\subsection{MUSIC Algorithm}

MUSIC algorithm, as a representative algorithm for modern spectral estimation, is based on the subspace theory in linear algebra. For an array composed of $N$ antenna elements, the covariance matrix $\mathbf{R}$ of the received signal can be decomposed into a set of orthogonal eigenvectors via eigenvalue decomposition (EVD). These eigenvectors span an $N$-dimensional signal space.

The key insight of the MUSIC algorithm is that this $N$-dimensional space can be further divided into two mutually orthogonal subspaces:

\begin{enumerate}
	\item Signal Subspace: This subspace is spanned by the eigenvectors corresponding to the largest eigenvalues of the covariance matrix. It contains the signal components and has a dimension equal to the number of signal sources, denoted as $K$.
	\item Noise Subspace: This subspace is spanned by the eigenvectors corresponding to the smallest eigenvalues. It contains only noise components and has a dimension of $N-K$.
\end{enumerate}

This orthogonality is the key to the super-resolution capability of the MUSIC algorithm. Specifically, for a steering vector $\mathbf{a}(\theta_k, r_k)$ that describes the true location of a target at $(\theta_k, r_k)$, it must satisfy:
\begin{equation}
	\mathbf{a}(\theta_k, r_k)^H \mathbf{U}_n = \mathbf{0}
\end{equation}
Here, $\mathbf{U}_n$ is the matrix composed of the eigenvectors of the noise subspace. Therefore, we can construct a pseudo-spectrum function and search over all possible $(\theta, r)$. When the test point $(\theta, r)$ exactly corresponds to the true target location, its steering vector is orthogonal to the noise subspace, and a sharp peak with infinite value will appear in the pseudo-spectrum function.

As mentioned earlier, the key to applying the MUSIC algorithm in the near-field lies in using the correct near-field steering vector. This steering vector must accurately characterize the phase differences produced at each array element when a spherical wave impinges on the linear array. According to the Fresnel approximation, it can be expressed as
	\begin{equation}
		\left[\mathbf{a}(\theta,r,f_{m})\right]_n = \left[e^{-j\frac{2\pi f_{m}}{c}\left(r-nd\sin\theta+\frac{n^{2}d^{2}\cos^{2}\theta}{2r}\right)}\right]
	\end{equation}

This steering vector is a nonlinear function of both angle and distance. Compared with the far-field steering vector, it incorporates the distance dimension through the quadratic phase term $\tfrac{n^2 d^2 \cos^2\theta}{2r}$, which provides the physical foundation for achieving super-resolution estimation in the distance domain.

\subsection{Fine Localization via MUSIC Algorithm}

\subsubsection{Single-Target Localization}
After obtaining the coarse target location $(\hat{r}_{k}^{(0)}, \hat{\theta}_{k}^{(0)})$ using the power spectrum peak method, we proceed to the second stage of the “Coarse-to-Fine” framework. The core task of this stage is to perform a high-precision two-dimensional spectral peak search within a very small local region determined by the coarse estimate, by applying the near-field MUSIC  subspace algorithm with super-resolution capability, thereby jointly resolving the final accurate position of the target.

Although the MUSIC algorithm is powerful, it requires spectral peak searching over the entire parameter space. For the two-dimensional joint estimation of $(\theta, r)$, such a global search leads to prohibitive computational complexity. The essence of the proposed framework is to leverage the coarse estimation result to restrict the MUSIC search region from a broad global domain to a very small local neighborhood centered at $(\hat{r}_{k}^{(0)}, \hat{\theta}_{k}^{(0)})$. This “focused regional probing” strategy makes the computational use of the MUSIC algorithm feasible, achieving an optimal balance between performance and complexity.

To further improve localization accuracy, local narrowband MUSIC refinement is performed over $S$ subcarriers around $m^\star$ (e.g., $m^\star - \Delta m, \ldots, m^\star + \Delta m$). For subspace analysis, we first need to estimate the covariance matrix of the received signal. For a subcarrier $f_m$ near the coarse peak index $m^\star$, the echo signal received at the BS, denoted as $\mathbf{y}_m \in \mathbb{C}^{N \times 1}$, is regarded as a “spatial snapshot.” Its sample covariance matrix is given as
	\begin{equation}
		\hat{\mathbf{R}}_m = \mathbf{y}_m \mathbf{y}_m^H
		\label{eq:sample_cov_matrix}
	\end{equation}

The covariance matrix $\hat{\mathbf{R}}_m$ obtained from a single snapshot is rank-one and thus cannot be directly applied to the MUSIC algorithm. To address this issue, we introduce the technique of Spatial Smoothing. This method divides the entire array into $P$ overlapping subarrays of size $M_s$, and averages the covariance matrices of these subarrays to restore the rank of the covariance matrix. The spatially smoothed covariance matrix $\bar{\mathbf{R}}_m$ is given as
\begin{equation}
	\bar{\mathbf{R}}_m = \frac{1}{P} \sum_{p=1}^{P} \mathbf{y}_m^{(p)} (\mathbf{y}_m^{(p)})^H
	\label{eq:spatial_smoothing}
\end{equation}
where $\mathbf{y}_m^{(p)}$ denotes the received vector of the $p$-th subarray.

After obtaining the full-rank covariance matrix $\bar{\mathbf{R}}_m$, eigenvalue decomposition (EVD) is performed as
\begin{equation}
	\bar{\mathbf{R}}_m = \mathbf{E} \mathbf{\Lambda} \mathbf{E}^H = \sum_{i=1}^{M_s} \lambda_i \mathbf{e}_i \mathbf{e}_i^H
\end{equation}

In the single-target scenario ($K=1$), the eigenvectors corresponding to the $M_s-1$ smallest eigenvalues, denoted as ${\mathbf{e}_2, \dots, \mathbf{e}_{M_s}}$, are used to construct the noise subspace matrix $\mathbf{U}_n = [\mathbf{e}_2, \dots, \mathbf{e}_{M_s}]$ .

The near-field MUSIC spatial spectrum function is defined as
\begin{equation}
		P_{\text{MUSIC},m}(\theta, r) = \frac{1}{\mathbf{a}_{M_s}(\theta, r, f_m)^H \mathbf{U}_n \mathbf{U}_n^H \mathbf{a}_{M_s}(\theta, r, f_m)}
		\label{eq:music_spectrum_detail}
\end{equation}

Here, $\mathbf{a}_{M_s}$ denotes the steering vector corresponding to the subarray dimension $M_s$. The spectrum function is evaluated only within the small local grid determined by the coarse estimate $(\hat{\theta}_{k}^{(0)} , \hat{r}_{k}^{(0)})$, in order to identify its peak.

Since the search is restricted to a small neighborhood around $m^\star$, the search region can be defined as
\begin{align}
	\theta &\in [\hat{\theta}_k^{(0)} - \Delta\theta, \hat{\theta}_k^{(0)} + \Delta\theta], \\
	r &\in [\hat{r}_k^{(0)} - \Delta r, \hat{r}_k^{(0)} + \Delta r],
\end{align}
Here, $\Delta \theta$ and $\Delta r$ can be set according to the accuracy of the coarse estimation (e.g., $\Delta \theta = 1^\circ$, $\Delta r = 1$ m).

To enhance the robustness of the algorithm in noisy environments, multiple subcarriers around the coarse peak index $m^\star$ are selected, i.e., $m \in \mathcal{S}$, and their local MUSIC spectra are fused through geometric averaging as
\begin{equation}
	P_{\text{GMUSIC}}(\theta,r) = \left(\prod_{m\in\mathcal{S}} P_{\text{MUSIC},m}(\theta,r)\right)^{\frac{1}{|\mathcal{S}|}}
\end{equation}

The geometric averaging method can effectively suppress spurious peaks that may exist in individual subcarriers, thereby improving the reliability of the spectrum function.The final high-accuracy position estimate is determined by the peak of the fused spectrum function as
\begin{equation}
	(\hat{\theta}_{k},\hat{r}_{k}) = \arg\max_{\theta,r} P_{\text{GMUSIC}}(\theta,r)
\end{equation}

The uniqueness and correctness of the near-field MUSIC solution are rigorously proven in Appendix \ref{appendix:music}.

Compared with the conventional global two-dimensional grid search–based MUSIC method, the proposed scheme significantly reduces computational complexity through the “Coarse-to-Fine” strategy. Meanwhile, the JAD trajectory design guarantees a unique mapping between frequency and space, thereby avoiding multi-frequency ambiguity and error accumulation.

\begin{figure}[!t]
\centering
\includegraphics[width=3in]{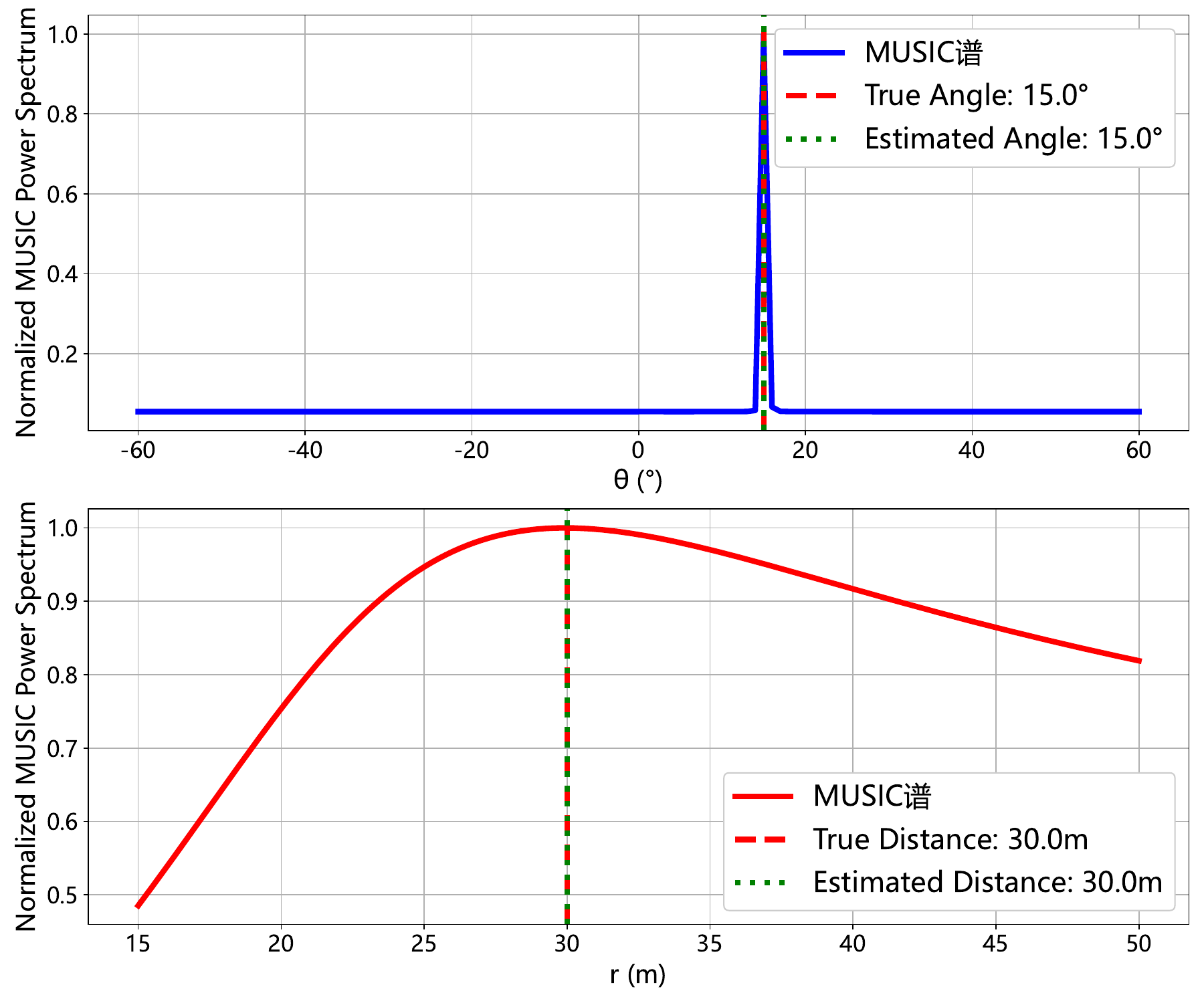}
\caption{Example of the two dimensional near-field MUSIC spatial spectrum in a single target scenario. A sharp and dominant peak appears at the true target location. }
\label{music}
\end{figure}

Fig. \ref{music} illustrates an example of the spatial spectrum obtained by the near-field MUSIC algorithm in a single-target scenario. It can be observed that the spectrum exhibits a prominent peak at the true target location, while no comparable peaks appear elsewhere. The designed JAD trajectory starts at $(-60^\circ, 15\ \text{m})$ and ends at $(60^\circ, 50\ \text{m})$, with the UE located at $(15^\circ, 30\ \text{m})$, as shown in Fig. 3. The coarse estimation provides an initial position of $(14.8^\circ, 29.5\ \text{m})$, around which the fine search is performed. To demonstrate the effectiveness of our method, the spectrum function is plotted over the entire region. As seen, the spectrum produces a distinct peak precisely at the true target location, with no other peaks of similar magnitude. This confirms that the near-field MUSIC algorithm can accurately localize the target.

Moreover, this result strongly validates that by performing eigenvalue decomposition (EVD) of the received signal covariance matrix and constructing the noise subspace, the proposed algorithm effectively exploits the spherical wavefront characteristics of near-field signals to precisely determine the target location. The sharpness of the peak further demonstrates the excellent angular and range resolution of the algorithm, with the main lobe being significantly higher than all sidelobes. This property is essential for high-accuracy localization in practice and provides solid verification of our theoretical derivations.

To clearly illustrate the proposed framework, we summarize the complete procedure in \textbf{Algorithm} \ref{alg:comprehensive_algorithm}, which integrates the efficiency of coarse estimation with the super-resolution capability of MUSIC-based fine estimation.

\begin{algorithm}[!t]
\caption{Proposed Coarse-to-Fine Single Scan Joint Localization Algorithm}
\label{alg:comprehensive_algorithm}
\begin{algorithmic}[1]
\STATE \textbf{Input:} Received echo signals of all subcarriers $\mathbf{Y} = [\mathbf{y}_0, \dots, \mathbf{y}_M]$.
\STATE \textbf{Output:} Final high-accuracy position estimate $(\hat{\theta}_k, \hat{r}_k)$.
\STATE \textbf{// Stage \Rmnum{1}:} Coarse Localization Stage
\STATE Compute the received power of each subcarrier $P_m = |\mathbf{y}_m|^2$, and find the peak index $m^* = \arg\max_m P_m$.
\STATE Obtain the coarse position estimate $(\hat{\theta}_{k}^{(0)}, \hat{r}_{k}^{(0)})$ by inverting the trajectory formulas.
\STATE \textbf{// Stage \Rmnum{2}:} Fine Localization Stage
\STATE Select the subcarrier set $\mathcal{S}$ around $m^*$.
\FOR{each subcarrier $m$ in $\mathcal{S}$}
\STATE Perform spatial smoothing on the received vector $\mathbf{y}_m$ to obtain $\bar{\mathbf{R}}m$.
\STATE Apply EVD to $\bar{\mathbf{R}}m$ to estimate the noise subspace $\mathbf{U}n$..\vspace{-0.9\baselineskip}
\STATE Evaluate $P_{\text{MUSIC},m}(\theta, r)$ over the local grid $\mathcal{G}$ centered at $(\hat{\theta}_{k}^{(0)}, \hat{r}_{k}^{(0)})$.
\ENDFOR
\STATE \textbf{// Stage \Rmnum{3}:} Spectrum Fusion and Output
\STATE Fuse all $P_{\text{MUSIC},m}(\theta, r)$ using geometric averaging to obtain $P_{\text{GMUSIC}}(\theta, r)$.
\STATE Identify the peak of the fused spectrum to yield the final estimate $(\hat{\theta}_k, \hat{r}_k) = \arg\max_{(\theta, r) \in \mathcal{G}} P_{\text{GMUSIC}}(\theta, r)$.
\RETURN $(\hat{\theta}_k, \hat{r}_k)$.

\end{algorithmic}
\end{algorithm}

\begin{figure}[!t]
\centering
\includegraphics[width=3in]{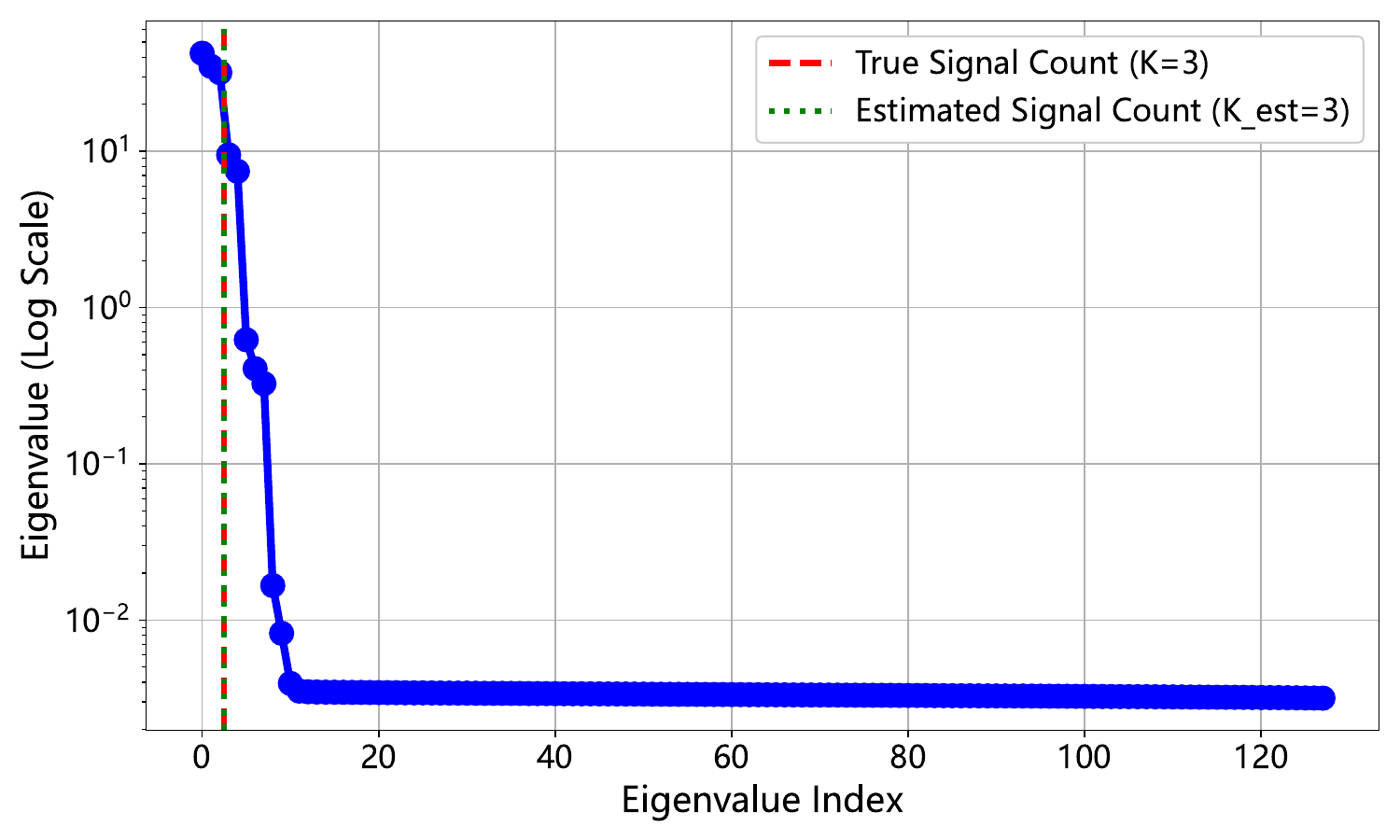}
\caption{Multi-User detection schematic diagram. The number of peaks identified from the eigenvalue spectrum of the received signal directly indicates the number of detected users.}
\label{tzz}
\end{figure}

\begin{figure*}[!t]
\centering
\subfloat[]{\includegraphics[width=2.5in]{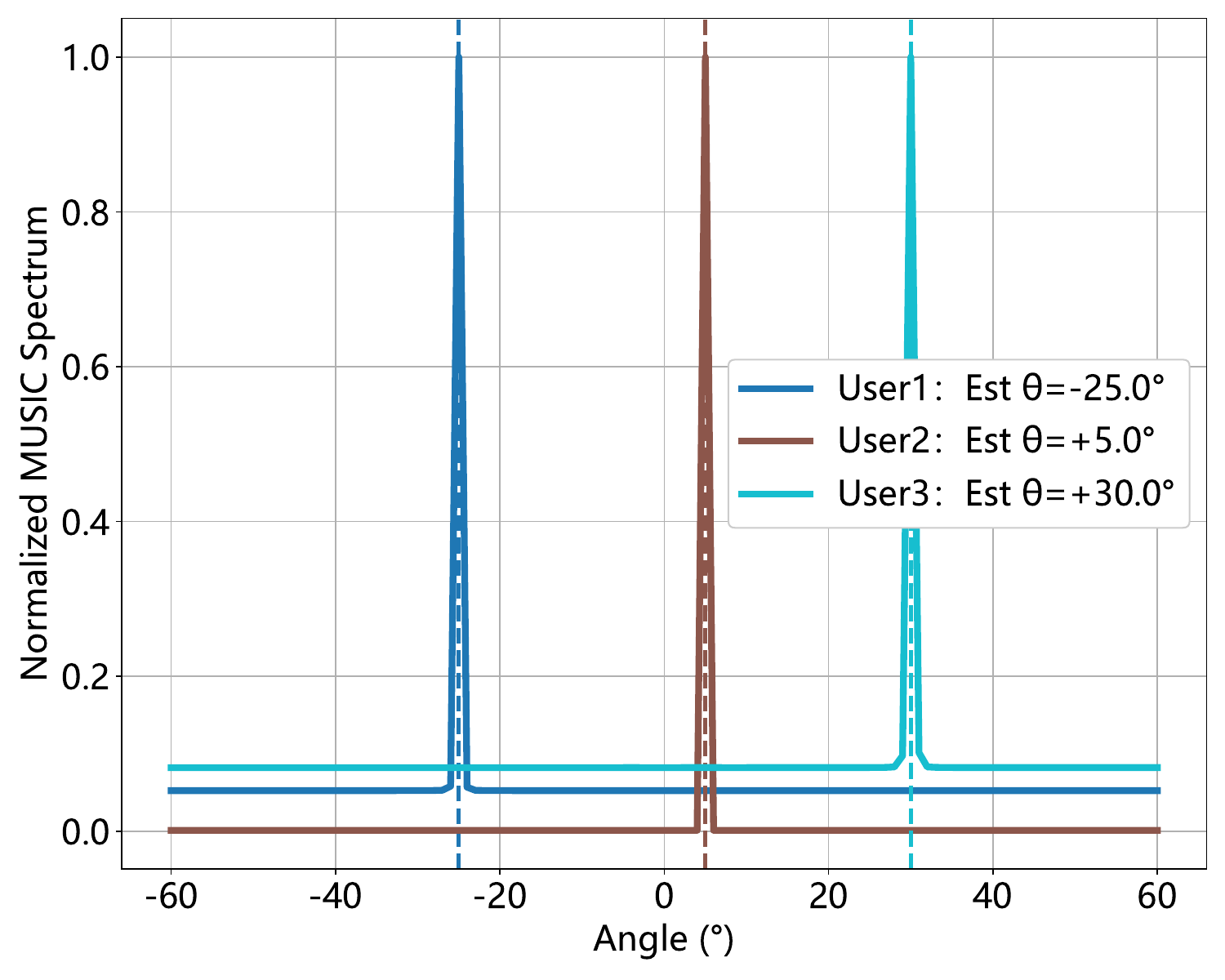}%
\label{Angle}}
\hfil
\subfloat[]{\includegraphics[width=2.5in]{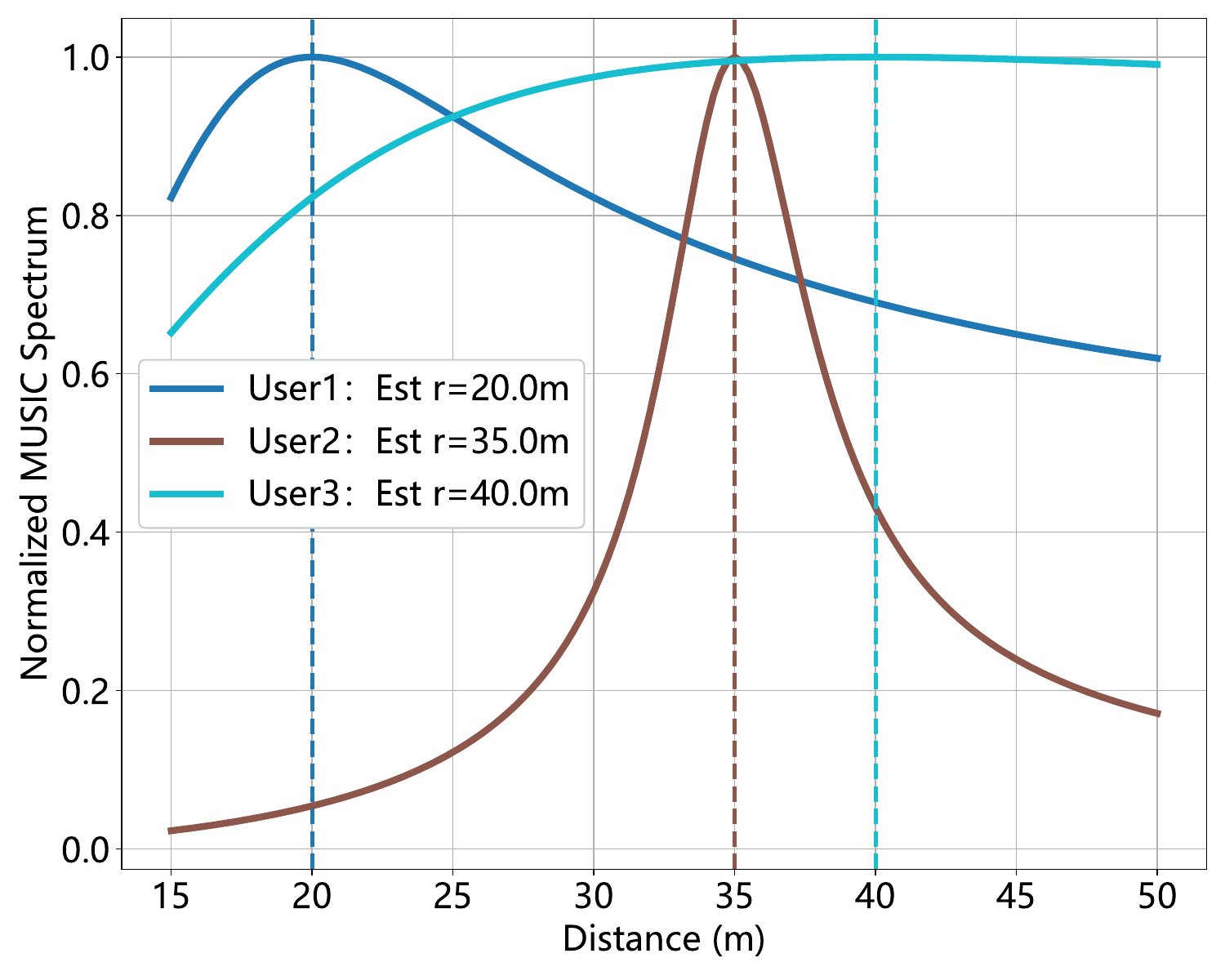}%
\label{distance}}
\caption{Joint localization results in Multi-User scenario. (a) Normalized MUSIC power spectrum in angular dimension. (b) Normalized MUSIC power spectrum in range dimension.}
\label{users}
\end{figure*}

\subsubsection{Multi-Target Joint Localization}

In practical communication and sensing scenarios, the BS often needs to simultaneously serve and sense multiple users, making multi-target joint localization a critical metric for evaluating the practicality of any algorithm. As shown in Fig. \ref{tzz}, the number of users can be directly identified from the eigenvalue curves. With the aid of the designed JAD trajectory, each user corresponds to a unique peak subcarrier index $m_k^\star$ in the frequency domain, thereby enabling user separation and coarse localization in the first stage.

In the refinement stage, the near-field MUSIC algorithm constructs the noise subspace and leverages its orthogonality to the true steering vectors, allowing the spectral peaks corresponding to different users to be clearly separated in the two-dimensional angle–distance domain. Thanks to its super-resolution property, MUSIC can effectively suppress mutual interference and maintain high resolution even when users are closely spaced in angle and distance.

Fig. \ref{users} illustrates the joint localization results in a multi-user scenario. It can be observed that sharp and prominent peaks appear precisely at the true locations of all users, while no spurious peaks emerge elsewhere. This result demonstrates that the proposed method not only achieves extremely high accuracy in single-user cases but also delivers outstanding localization performance in multi-user environments.

It is worth emphasizing that, compared to conventional methods relying on power peaks or multi-stage estimation, the proposed “JAD trajectory and coarse-to-fine two stage” framework enables simultaneous acquisition of angle and distance information for multiple users within a single scan, thereby avoiding error propagation and delay accumulation. Moreover, the multi-subcarrier fusion strategy significantly enhances the robustness of the spectrum in multi-target environments, ensuring that the algorithm maintains high resolution and accuracy even under low SNR conditions or when the number of users increases.

\begin{figure*}[!t]
\centering
\subfloat[]{\includegraphics[width=2.5in]{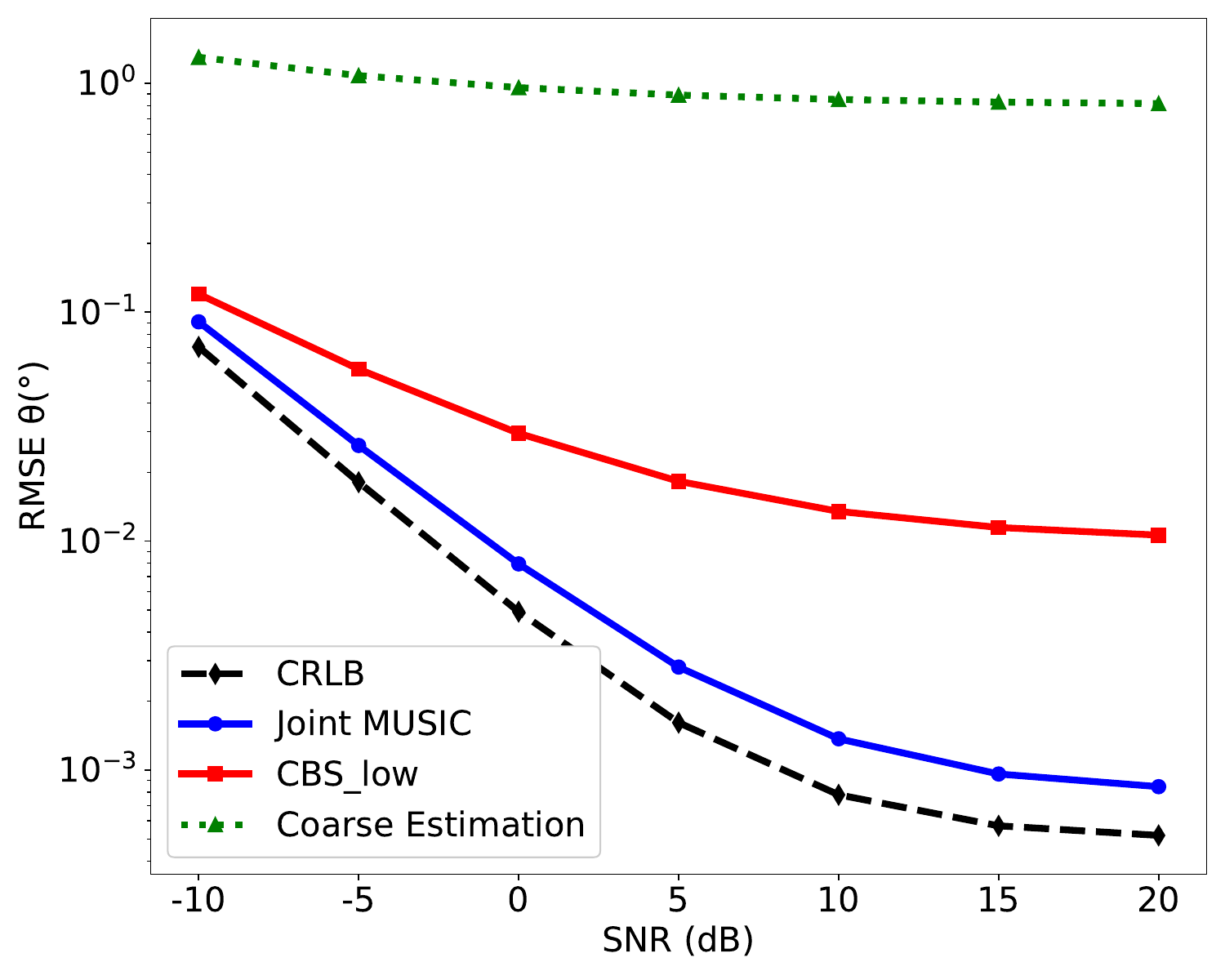}%
\label{angle_rmse}}
\hfil
\subfloat[]{\includegraphics[width=2.5in]{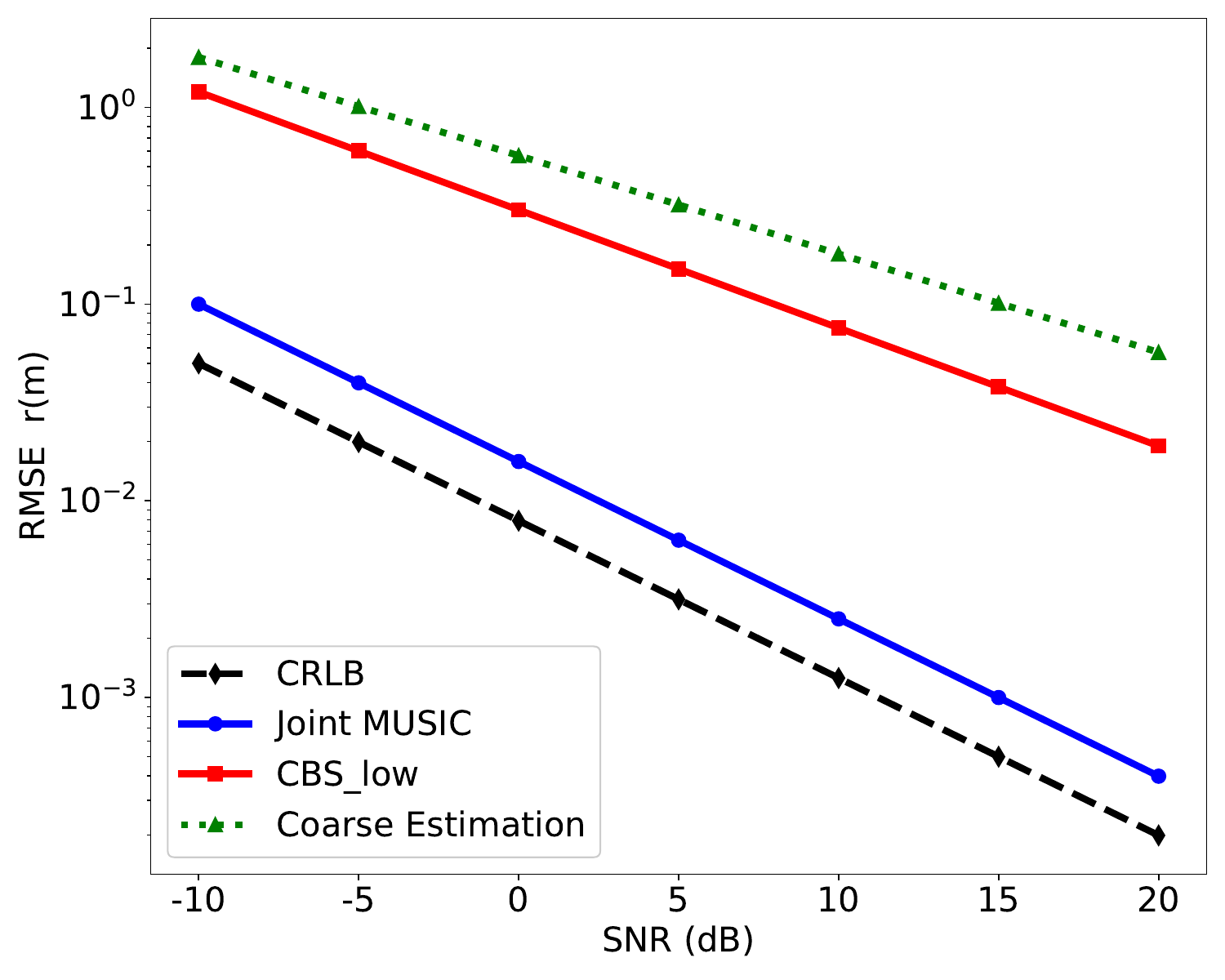}%
\label{distance_rmse}}
\caption{RMSE performance comparison versus SNR: (a) angle RMSE and (b) range RMSE.}
\label{rmse_single}
\end{figure*}

\begin{figure*}[!t]
\centering
\subfloat[]{\includegraphics[width=2.5in]{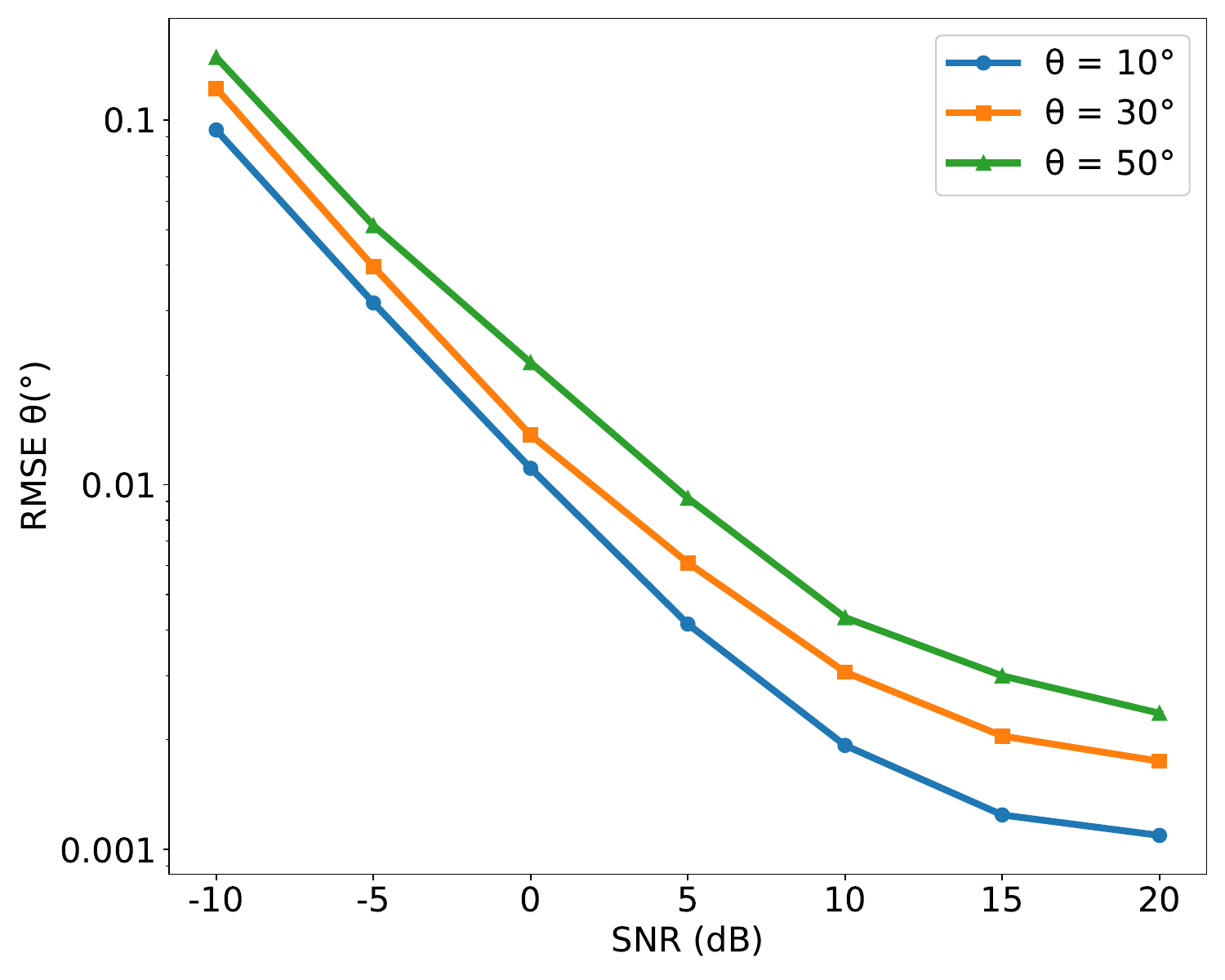}%
\label{angless}}
\hfil
\subfloat[]{\includegraphics[width=2.5in]{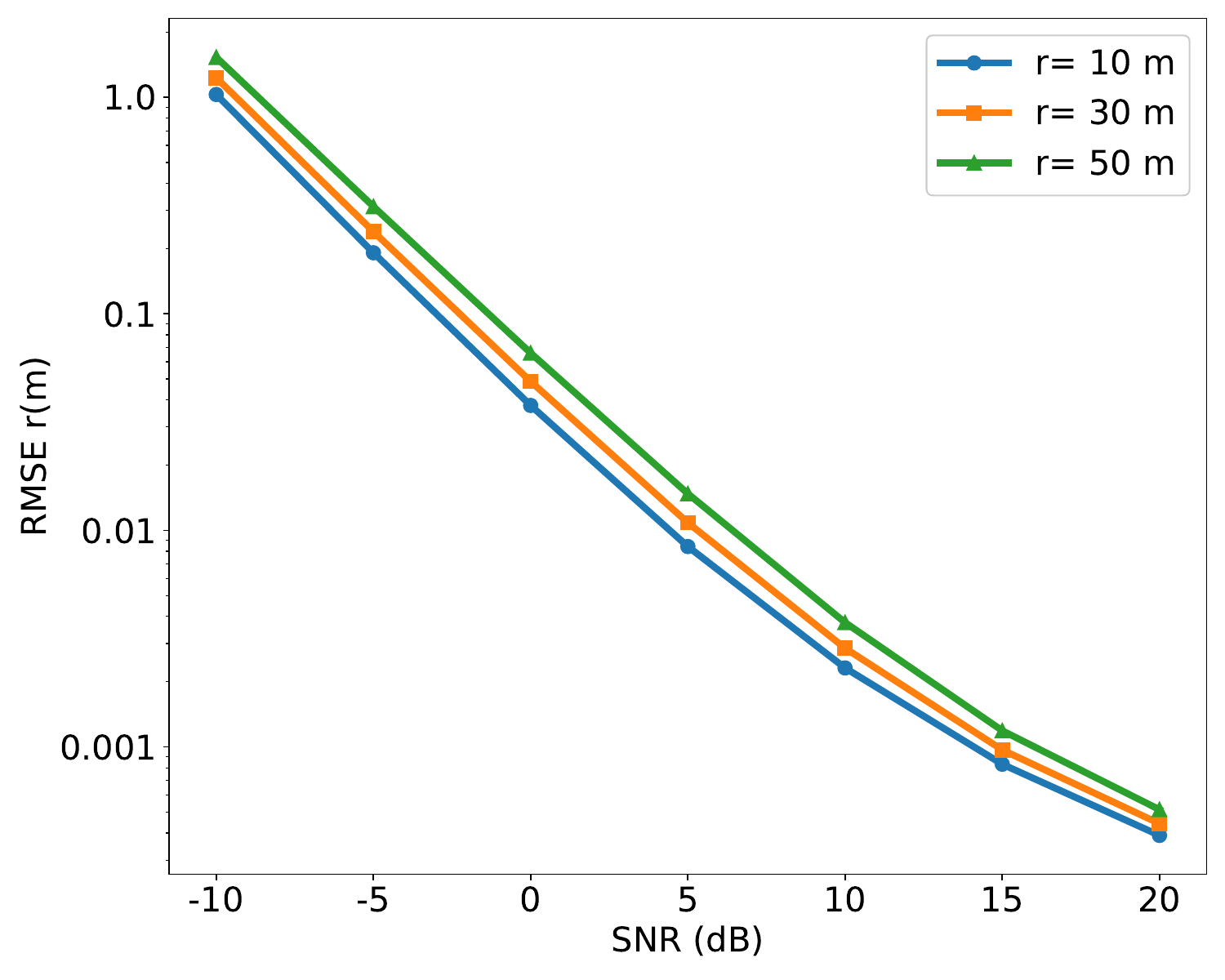}%
\label{diatancess}}
\caption{Robustness analysis under different geometric configurations: (a) angle RMSE for incident angles $\theta = 10^\circ, 30^\circ, 50^\circ$; (b) range RMSE for target distances $r = 10$ m, $30$ m, and $50$ m. }
\label{rmse}
\end{figure*}

\section{Simulation Results}
\label{sec:simulation}
In this section, simulation results are presented to validate the effectiveness of the proposed beam-squint–assisted joint angle–distance localization scheme. We consider a wideband XL-MIMO system operating at a carrier frequency of $f_c = 60$ GHz, where the BS is equipped with a uniform linear array (ULA) consisting of $N_t = 256$ antennas and employs a true-time-delay (TTD) aided beamforming architecture. The system bandwidth is set to $B = 3$ GHz, which is equally divided into $M = 2048$ subcarriers. Unless otherwise specified, the inter-element spacing of the ULA is $d = \lambda/2$, and the noise is modeled as complex Gaussian with variance $\sigma^2$. Consider the Electromagnetic and Rayleigh distance ,the sensing range of BS system is set to  $\{(\theta, r) \mid  -60^\circ \leq \theta \leq 60^\circ, 15\ \text{m} \leq r \leq 50\ \text{m}\}$. Both single-user and multi-user near-field scenarios are considered to comprehensively evaluate the proposed scheme in terms of localization accuracy, robustness, and computational efficiency.

For performance comparison, we consider the following schemes.
\begin{itemize}
	\item \textbf{Two-Step Localization (CBS-Low):} This method originates from the pioneering work of Luo et al. in near-field ISAC systems and represents an advanced engineering approach that also leverages controllable beam squint for localization. Unlike our proposed joint estimation scheme, CBS-Low adopts a decoupled two stage estimation strategy: first estimating the angle and then the distance. This sequential approach may lead to error propagation and accumulation, thereby limiting its localization accuracy\cite{luo2023beam}.
	\item \textbf{Coarse Estimation based on Power Peak :} This baseline is an extremely low-complexity method whose principle is identical to the coarse stage of our proposed scheme. Specifically, the algorithm executes a single JAD trajectory scan, computes the received power across all subcarriers, identifies the peak index $m^*$, and directly maps it to a spatial coordinate $(\hat{r}_{k}^{(0)}, \hat{\theta}_{k}^{(0)})$ as the final estimate. By including this baseline, we aim to quantitatively demonstrate the significant performance gain achieved by incorporating the MUSIC-based fine estimation stage.
	\item \textbf{Cramér–Rao Lower Bound (CRLB):} The CRLB is not an algorithm but rather a theoretical bound that characterizes the minimum possible RMSE achievable by any unbiased estimator under the given system model and Gaussian noise assumption. In this work, the CRLB is derived from the Fisher Information Matrix (FIM) constructed under the near-field sensing model, with the inverse matrix providing the theoretical error floor. By comparing the RMSE of our proposed algorithm with the CRLB, we can assess its effectiveness and statistical efficiency, i.e., how closely it approaches the "theoretical optimum". The detailed derivation is provided in Appendix \ref{appendix:crlb}.
\end{itemize}

\subsection{simulation for single-user scenario}

Fig. \ref{rmse_single} \subref{angle_rmse}reports the angle estimation RMSE as a function of SNR. The proposed beam-squint–assisted near-field MUSIC joint estimator outperforms both benchmark methods—the cascaded beam-squint two-step scheme (CBS-Low) and the power peak coarse estimator—over the entire SNR range. In particular, our method achieves substantially lower variance at low-to-moderate SNR and approaches the CRLB in the moderate-to-high SNR regime. This performance gain is explained by three key factors: (1) the JAD trajectory establishes a unique frequency–space mapping that enables reliable separation of angular candidates in the coarse stage; (2) the localized near-field MUSIC exploits array phase information and spatial smoothing to achieve super-resolution, thereby dramatically reducing angular estimation error; and (3) geometric averaging across multiple subcarriers suppresses spurious peaks arising from single-subcarrier noise, stabilizing peak detection. By contrast, CBS-Low suffers from error propagation from the first (angle) stage, and the power-peak baseline is limited by spatial quantization and noise sensitivity, preventing it from approaching the CRLB.

Fig. \ref{rmse_single} \subref{distance_rmse} shows the range estimation RMSE versus SNR. The proposed scheme yields a clear advantage in range estimation as well: it outperforms both the two-step CBS-Low and the power peak baseline at low SNR and approaches the CRLB at higher SNR. Range estimation is highly sensitive to angular errors, which explains why the cascaded two step method exhibits degraded range performance due to error propagation from the angle stage. In contrast, our single scan joint framework avoids such propagation; the near-field MUSIC directly exploits the distance dependent quadratic phase term in the steering vector to jointly resolve angle and range peaks in the 2-D spectrum, leading to superior range resolution across SNR regimes. Moreover, multi-subcarrier fusion and spatial smoothing improve robustness to noise and enable effective cooperation among frequency-dependent focal points under beam-squint, further enhancing range accuracy.

\subsection{simulation for multi-user scenario}

Fig. \ref{rmse} \subref{angless} illustrates the angle RMSE as a function of SNR for different UE angles, i.e., $\theta = 10^\circ, 30^\circ, 50^\circ$. The proposed joint localization algorithm consistently achieves high accuracy and stability across all tested angular configurations, approaching the CRLB in the moderate-to-high SNR regime. This confirms that the designed JAD trajectory effectively covers a wide angular range and ensures robust angular resolution under diverse geometric conditions. By contrast, CBS-Low exhibits more pronounced errors at larger incident angles due to the error propagation from its cascaded structure, while the power-peak baseline becomes highly unstable in the low-SNR regime owing to its noise sensitivity. Overall, the proposed method maintains its super-resolution advantage under angular diversity, highlighting its applicability to multi-user scenarios.

Fig. \ref{rmse} \subref{diatancess} presents the range RMSE versus SNR for different target distances, i.e., $r = 10$ m, $30$ m, and $50$ m. The proposed scheme maintains stable performance across all distance configurations and approaches the CRLB in the moderate-to-high SNR regime. This demonstrates that the near-field steering vector with its quadratic phase term enables accurate range resolution under varying propagation distances. In comparison, CBS-Low suffers more severe errors at longer ranges, reflecting the limitations of its two-step estimation in large-aperture near-field scenarios, while the power-peak baseline consistently yields larger errors and fails to adapt to far-distance cases. Overall, the proposed algorithm preserves high accuracy and robustness under both distance and SNR variations, confirming its scalability for practical multi-user communication and sensing systems.s
\section{Conclusion}
\label{sec:conclusion}

In this paper, we have investigated the problem of high-accuracy user localization in near-field wideband XL-MIMO systems. We first analyzed the near-field beam-squint effect and designed a joint angle-distance (JAD) trajectory that effectively covers the two-dimensional sensing region. Then, we proposed a novel “Coarse-to-Fine” two stage localization framework. The first stage performs fast coarse localization by analyzing the power spectrum of the echo signal, while the second stage conducts a high-precision two-dimensional spectral peak search using a near-field MUSIC subspace algorithm with super-resolution capabilities. This approach enables joint estimation of both angle and distance in a single step, avoiding error propagation and delay accumulation. Simulation results demonstrated that the proposed method achieves extremely high accuracy in both single-user and multi-user scenarios, validating its effectiveness and practicality.

{\appendices

\section{}
\label{appendix:music}
This section aims to provide a theoretical proof that, for the proposed MUSIC algorithm based on near-field steering vectors, the peak of the spatial spectrum is unique, and the spatial coordinate $(\theta, r)$ corresponding to this peak precisely represents the true position of the target.

The proof consists of two steps: first, we show that at the true target location, the spectrum function exhibits a peak (correctness); second, we demonstrate that at any non-true location, no peak occurs (uniqueness).

\subsection{Proof of  correctness}

According to the monostatic sensing model, the ideal signal covariance matrix is given by $R = P_s a(\theta_k, r_k) a(\theta_k, r_k)^H + \sigma^2_n I$.where $a(\theta_k, r_k)$ denotes the steering vector. By performing eigenvalue decomposition (EVD) on $R$, the signal subspace $\mathcal{U}_s$ is spanned by the true steering vector $a(\theta_k, r_k)$, while the noise subspace $\mathcal{U}_n$ is orthogonal to $\mathcal{U}_s$.

This implies that the true steering vector $a(\theta_k, r_k)$ must be orthogonal to any basis vector of the noise subspace $\mathcal{U}_n$. Let $U_n$ denote the matrix composed of all basis vectors of $\mathcal{U}_n$, then it follows that
\begin{equation}
	a(\theta_k, r_k)^H U_n = \mathbf{0}
	\label{eq:orthogonality_proof}
\end{equation}

The MUSIC pseudo-spectrum function is defined as
\begin{equation}
	P_{\text{MUSIC}}(\theta, r) = \frac{1}{a(\theta, r)^H U_n U_n^H a(\theta, r)}
	\label{eq:music_spectrum_proof_app}
\end{equation}

When the true target location $(\theta_k, r_k)$ is substituted into the above expression, the denominator becomes zero according to \eqref{eq:orthogonality_proof}.

In the ideal case, this leads to $P_{\text{MUSIC}}(\theta_k, r_k) \to \infty$. This proves that at the true target location, the MUSIC spectrum function necessarily exhibits a peak, which validates the correctness of the solution.

\subsection{Proof of uniqueness}

To establish uniqueness, we must prove that for any “false” location $(\theta', r') \neq (\theta_k, r_k)$, the spectrum value $P_{\text{MUSIC}}(\theta', r')$ cannot form a peak, i.e., its denominator is strictly non-zero.

We proceed by contradiction. Suppose that a false location $(\theta', r')$ also yields a peak. Then its denominator must also vanish, which implies that $a(\theta', r')$ is orthogonal to the noise subspace. In the single-target case, the signal subspace is one-dimensional, hence $a(\theta', r')$ must be linearly dependent on the true steering vector $a(\theta_k, r_k)$:
\begin{equation}
	a(\theta', r') = c \cdot a(\theta_k, r_k)
	\label{eq:ambiguity_condition_proof}
\end{equation}
where $c$ is a nonzero complex scalar.

This equivalently requires that their phase difference is constant across all antennas $n$.
\begin{equation}
	\Phi_n(\theta', r') - \Phi_n(\theta_k, r_k) = \text{const}, \quad \forall n
\end{equation}

By substituting the phase expressions and simplifying, we obtain

\begin{multline}
    \underbrace{(r_k - r')}_{C_0} + n_d \cdot \underbrace{d(\sin\theta' - \sin\theta_k)}_{C_1} \\
    + n_d^2 \cdot \underbrace{\frac{d^2}{2}\left(\frac{\cos^2\theta_k}{r_k} - \frac{\cos^2\theta'}{r'}\right)}_{C_2} = \text{const}'
    \label{eq:polynomial_proof}
\end{multline}

This expression forms a quadratic polynomial in the antenna index $n_d$. For this polynomial to remain constant across all $N$ distinct values of $n_d$, both the linear and quadratic coefficients must vanish simultaneously (for arrays with $N \ge 3$).

Therefore, the only condition under which \eqref{eq:polynomial_proof} holds is $(\theta', r') = (\theta_k, r_k)$. This contradicts our initial assumption that a false location $(\theta', r') \neq (\theta_k, r_k)$ could also generate a peak.

Thus, the assumption is invalid. No “ambiguous point” distinct from the true location can produce a steering vector that is linearly related to the true one. This implies that for any $(\theta', r') \neq (\theta_k, r_k)$, the steering vector $a(\theta', r')$ must be non-orthogonal to the noise subspace, and the denominator of the MUSIC spectrum remains strictly positive, preventing any peak formation.This completes the proof of the uniqueness of the solution.

\section{Proof of CRLB}
\label{appendix:crlb}
The Cramér–Rao lower bound (CRLB) provides a theoretical lower limit on the mean squared error of any unbiased estimator and serves as a fundamental benchmark for evaluating the statistical efficiency of the proposed estimation algorithm. In this section, we derive the CRLB for joint angle and distance estimation in a monostatic, single-target, near-field sensing scenario.

Assume that the base station is equipped with a uniform linear array (ULA) of $N$ elements and receives $L$ snapshots of echo signals from a single target located at $(\theta, r)$. At the $l$-th snapshot, the received signal vector $y_l \in \mathbb{C}^{N \times 1}$ can be modeled as
\begin{equation}
	y_l = a(\theta, r)s_l + n_l, \quad l=1, \dots, L
\end{equation}
where $s_l$ is the complex reflection coefficient of the $l$-th snapshot (accounting for path loss and target RCS), and $n_l \sim \mathcal{CN}(0, \sigma^2 I)$ denotes zero-mean circularly symmetric complex Gaussian noise. The vector $a(\theta, r)$ represents the near-field steering vector, whose $n$-th element is given as
\begin{align}
    [a(\theta, r)]_n &= \exp\left\{j \Phi_n(\theta, r)\right\} \notag \\
    &= \exp\left\{-j\frac{2\pi f_c}{c}\left(r - n_d d \sin\theta + \frac{(n_d d)^2 \cos^2\theta}{2r}\right)\right\}
\end{align}
where $n_d = n - (N-1)/2$ denoting the antenna index referenced to the array center.

The unknown parameters to be estimated are the angle $\theta$ and distance $r$. In practice, the reflection coefficients $s_l$ and noise variance $\sigma^2$ may also be unknown. For analytical tractability, we consider a more general case in which the complex amplitudes $s_l$ are treated as unknown deterministic parameters. The real-valued parameter vector to be estimated is then defined as $\boldsymbol{\eta} = [\theta, r]^T$.

For the above signal model, the Fisher information matrix (FIM) of the parameter vector $\boldsymbol{\eta}$, denoted by $J(\boldsymbol{\eta})$, is a $2 \times 2$ matrix.In the presence of complex Gaussian noise, its $(i,j)$-th entry can be computed as
\begin{equation}
		[J(\boldsymbol{\eta})]_{i,j} = \frac{2}{\sigma^2} \sum_{l=1}^{L} |s_l|^2 \cdot \text{Re} \left\{ \left(\frac{\partial a}{\partial \eta_i}\right)^H \left(\frac{\partial a}{\partial \eta_j}\right) \right\}
\end{equation}
where $\eta_1 = \theta$, $\eta_2 = r$, and $\tfrac{\partial a}{\partial \eta_i}$ denotes the partial derivative of the steering vector with respect to the parameter $\eta_i$.

First, we compute the partial derivatives of the steering vector with respect to $\theta$ and $r$.Let $d_\theta = \tfrac{\partial a(\theta, r)}{\partial \theta}$ and $d_r = \tfrac{\partial a(\theta, r)}{\partial r}$.Their $n$-th elements are given as
\begin{align}
		[d_\theta]_n &= [a(\theta, r)]_n \cdot j \frac{\partial \Phi_n(\theta, r)}{\partial \theta} \\
		[d_r]_n &= [a(\theta, r)]_n \cdot j \frac{\partial \Phi_n(\theta, r)}{\partial r}
\end{align}
where the partial derivatives of the phase are given as
\begin{align}
		\frac{\partial \Phi_n(\theta, r)}{\partial \theta} &= -\frac{2\pi f_c}{c} \left( -n_d d \cos\theta - \frac{(n_d d)^2 \cdot 2\cos\theta(-\sin\theta)}{2r} \right) \nonumber \\
		&= \frac{2\pi f_c}{c} \left( n_d d \cos\theta - \frac{(n_d d)^2 \sin(2\theta)}{2r} \right) \\
		\frac{\partial \Phi_n(\theta, r)}{\partial r} &= -\frac{2\pi f_c}{c} \left( 1 - \frac{(n_d d)^2 \cos^2\theta}{2r^2} \right)
\end{align}

The total signal-to-noise ratio (SNR) is defined as $\text{SNR} = \frac{L \sum|s_l|^2}{N\sigma^2}$.For simplicity, we consider the single-snapshot case $L=1$, and define the effective SNR with normalized signal power as $\text{SNR} = \frac{L \sum|s_l|^2}{N\sigma^2}$.The FIM can then be expressed as

\begin{multline}
    J = \frac{2|s|^2 L}{\sigma^2} \begin{pmatrix}
        \text{Re}\{d_\theta^H d_\theta\} & \text{Re}\{d_\theta^H d_r\} \\
        \text{Re}\{d_r^H d_\theta\} & \text{Re}\{d_r^H d_r\}
    \end{pmatrix} \\
    = 2 \text{SNR}_{\text{eff}} L \begin{pmatrix}
        \|d_\theta\|^2 & \text{Re}\{d_\theta^H d_r\} \\
        \text{Re}\{d_r^H d_\theta\} & \|d_r\|^2
    \end{pmatrix}
\end{multline}
where $|d_\theta|^2 = \sum_{n=1}^{N} \left|\tfrac{\partial \Phi_n}{\partial \theta}\right|^2$ and $|d_r|^2 = \sum_{n=1}^{N} \left|\tfrac{\partial \Phi_n}{\partial r}\right|^2$.

The CRLB is given by the inverse of the FIM. For a $2 \times 2$ matrix $J = \begin{pmatrix} A & B \\ B & C \end{pmatrix}$, the inverse is $J^{-1} = \frac{1}{AC-B^2} \begin{pmatrix} C & -B \\ -B & A \end{pmatrix}$.

Therefore, the CRLBs for angle and distance estimation are given as

\begin{align}
    \text{CRLB}(\theta) &= [J^{-1}]_{1,1} = \frac{J_{2,2}}{\det(J)} \notag \\  
    &\quad = \frac{\|d_r\|^2}{2 \text{SNR}_{\text{eff}} L \left( \|d_\theta\|^2 \|d_r\|^2 - (\text{Re}\{d_\theta^H d_r\})^2 \right)} \label{eq:crlb_theta} \\
    \text{CRLB}(r) &= [J^{-1}]_{2,2} = \frac{J_{1,1}}{\det(J)} \notag \\  
    &\quad = \frac{\|d_\theta\|^2}{2 \text{SNR}_{\text{eff}} L \left( \|d_\theta\|^2 \|d_r\|^2 - (\text{Re}\{d_\theta^H d_r\})^2 \right)} \label{eq:crlb_r}
\end{align}

The lower bound of the root-mean-square error (RMSE) is given by $\text{RMSE} \ge \sqrt{\text{CRLB}}$. Equations \eqref{eq:crlb_theta} and \eqref{eq:crlb_r} serve as the theoretical performance bounds adopted in the simulations of this paper.
}

\bibliographystyle{IEEEtran}
\bibliography{IEEEabrv,Beam_Squint_Assisted_Joint_Angle-Distance_Localization_for_Near-Field_Communications}

\vfill

\end{document}